\documentclass[3p]{elsarticle}
\journal{Mechanics of Materials}

\biboptions{sort&compress}

\usepackage{amsmath,amssymb,mathtools}
\usepackage{graphicx}
\usepackage{xcolor}
\usepackage{tabularx}
\usepackage{array}
\usepackage{bm}
\usepackage{hyperref}
\usepackage{makecell}
\usepackage{subcaption}
\usepackage{float}
\usepackage{lineno}
\usepackage{xcolor}
\usepackage{soul}

\definecolor{myblue}{named}{black}
\newcommand{\myblue}[1]{\textcolor{myblue}{#1}}

\hypersetup{colorlinks=true,citecolor=blue!60!black,linkcolor=blue!60!black}

\newcommand{\Fth}{\F^{\rm th}}

\newcommand{\Fz}{\F^{0}}
\newcommand{\Fez}{\F^{0, \rm e}}
\newcommand{\Pz}{\bfP^{0}}

\newcommand{\OmegaR}{\Omega_{\rm R}}

\newcommand{\aX}{a(\X)}
\newcommand{\DG}{\Delta G}

\newcommand{\Tm}{T_m}

\newcommand{\RefDom}{\Omega_{\rm R}}

\newcommand{\CurDom}{\Omega}

\newcommand{\SeedRef}{\Omega_i}

\newcommand{\xcur}{\bfx}

\newcommand{\chicur}{\boldsymbol\chi}

\newcommand{\Fpre}{\bfF^0}

\newcommand{\FpreE}{\bfF^{0, \rm e}}
\newcommand{\FpreTh}{\bfF^{0, \rm th}}
\newcommand{\Fetath}{\F^{\eta, \rm th}}
\newcommand{\Fetael}{\F^{\eta, \rm e}}

\DeclareMathOperator{\diag}{diag}

\newcommand{\bfA}{{\bf A}}%
\newcommand{\bfF}{{\bf F}}%
\newcommand{\bfP}{{\bf P}}%
\newcommand{\bft}{{\bf t}}%
\newcommand{\bfu}{{\bf u}}%
\newcommand{\bfx}{{\bf x}}\newcommand{\bfX}{{\bf X}}%
\newfont{\tenbfit}{cmmib10}%
\newfont{\svnbfit}{cmmib8}%
\newfont{\tenbfsl}{cmbxti10}
\newfont{\mmit}{cmmi10}
\newfont{\smit}{cmmi9}
\newfont{\bfMit}{cmmi5}
\newfont{\tenbbb}{msbm10}%
\newfont{\svnbbb}{msbm8}%
\newfont{\tenssit}{cmssqi8 at 10pt}%
\newfont{\svnssit}{cmssqi8 at 7pt}%
\newfont{\gothic}{eufm10}%
\newfont{\sgothic}{eufm7}%

\newcommand{\pards}[2]{\mbox{$\dfrac{\partial #1}{\partial {#2 }}$}}

\newcommand{\Blj}{\mbox{$\Big[\kern-0.275em\Big[$}}
\newcommand{\Brj}{\mbox{$\Big]\kern-0.275em\Big]$}}

\newcommand{\id}{{\bf 1}}
\newcommand{\X}{\bfX}

\newcommand{\F}{\bfF}

\newcommand{\tendot}{\mskip-3mu:\mskip-2mu}

\newcommand{\Fe}{\bfF^e}

\begin{document}

\begin{frontmatter}

\title{Energetics of Nucleation in Finitely Deformed, Phase-Transforming Soft Solids}

\author[unh]{Mrityunjay Kothari}
\address[unh]{Department of Mechanical Engineering, University of New Hampshire, Durham, NH 03824, USA}

\begin{abstract}
Classical nucleation theory describes the rate at which stable nuclei form within a metastable parent phase by crossing a free-energy barrier set by competing bulk and interfacial energies. 
In an elastic material, a pre-existing stress state modifies this barrier through an elastic  contribution to the bulk driving force. 
This contribution is well characterized for linear elastic materials, but the corresponding finite-deformation result for soft solids remains less developed. 
The gap is computationally significant: in simulations that sample candidate nuclei throughout a stressed body, direct evaluation of the elastic contribution to free-energy change would require solving a new nonlinear elasticity boundary-value problem for each possible nucleus.
Here, we derive an asymptotic expansion of the equilibrium elastic potential energy change for a hyperelastic body before and after formation of a small transformed region. 
The expansion is with respect to the amplitude of an isotropic transformation strain, while the pre-existing deformation and stress may be finite. 
At leading order, the elastic contribution to the formation energy is determined entirely by the known untransformed equilibrium fields, with additional terms accounting for stiffness contrast between the parent and transformed phases. 
Incorporating this into classical nucleation theory yields the stress-shifted transformation temperature, critical radius, and nucleation barrier.
Representative results are shown for a compressible neo-Hookean solid under hydrostatic, uniaxial, and equibiaxial loading; tensile stresses promote nucleation and compressive stresses suppress it when transformation strain is expansive.
Comparison with the corresponding linear-elastic result shows that finite-deformation effects can substantially change the predicted energy barrier at moderate stretches. 
\end{abstract}

\end{frontmatter}

\section{Introduction}
Nucleation of transformed domains in soft solids arises in a variety of settings, such as freezing of tissues in cryopreservation and strain-induced crystallization of elastomers \cite{xu2021concepts, hartquist2023elastomer, wang2024molecular, chen2023cryopreservation,saeedi2026thermo}. 
In such phenomena, the energetic cost of a nucleus is not determined solely by local thermal, chemical, and interfacial driving forces; it also depends on the pre-existing deformation of the surrounding matrix, on the transformation strain of the nucleus, and on the contrast of elastic properties between the nucleus and the surrounding matrix.
These factors add a bulk elastic potential energy contribution to the free-energy change for nucleation \cite{russell1980nucleation, cahn1962coherent, khachaturyan2013theory}.
This is the origin of stress effects on material microstructure and has been explored in considerable depth for linear elastic materials, thanks to pioneering work by J. D. Eshelby \cite{eshelby1957determination, mura2013micromechanics, lu1997martensitic, lue2000micro, shen2006effect}.
However, for soft solids, which show finite deformations and non-linear elasticity, the question is less settled: how does a pre-existing finite-deformation stress state alter the thermodynamic barrier for nucleation?

Classical nucleation theory (CNT) gives the free-energy change for a spherical nucleus of radius $R$ as
\begin{equation}
  \DG(R)=4\pi R^2\gamma-\frac{4\pi}{3}R^3\Delta f,
  \label{eq:intro_CNT}
\end{equation}
where $\gamma$ is the interfacial energy and $\Delta f$ is the bulk driving force per unit volume~\cite{gibbs1878equilibrium, volmer1926nucleus, becker1935kinetische, turnbull1949formation}.  
For specificity, we will consider a thermally triggered phase transformation for which $\Delta f$ can be linearized near the transformation temperature.
For a transformed phase that is favored below $\Tm$, we write
\begin{equation}
	\Delta f = \frac{L(\Tm-T)}{\Tm},
\end{equation}
where $T$ is the current temperature, $\Tm$ is the transformation temperature in the absence of elastic effects, and $L$ is the latent heat per unit volume\footnote{\myblue{We do not consider any direct dependence of latent heat $L$ on stress in this study. Should such an effect be important for a material, the current theory can be extended in accordance with equation (2.26) in Saeedi et al. \cite{saeedi2026thermo}.}}.

It is known that hydrostatic pressure shifts the melting temperature through the Clausius--Clapeyron relation~\cite{clapeyron1834, callen2006thermodynamics}, but soft solids can sustain heterogeneous, thermoelastic stress states for which a scalar pressure correction is not sufficient.  
The finite-deformation problem is not solved by replacing pressure with an ad hoc stress measure: the elastic contribution must be obtained from the work-conjugate pair that appears in the equilibrium energy difference between the untransformed and transformed systems.  

The need for such a result is especially clear in simulations of nucleation in stressed soft solids. 
\myblue{In stochastic or phase-field-type simulations of nucleation, candidate nuclei may be sampled throughout a stressed body and assigned nucleation probabilities from local free-energy barriers \cite{simmons2000phase, shen2007effect, heo2010incorporating, simmons2004microstructural}}. 
A direct finite-deformation treatment would require solving a separate nonlinear elasticity boundary-value problem for every possible candidate nucleus in order to compute its energy barrier. 
Such a procedure is generally prohibitive. 
The aim of the present work is to obtain an asymptotic energetic contribution that can be evaluated from the pre-existing equilibrium stress and deformation fields, thereby converting the elastic part of the nucleation barrier into a local post-processing calculation, within the range of validity of the small-transformation-strain expansion.

This paper derives the elastic contribution directly from the equilibrium potential energy of a hyperelastic body before and after formation of a small transformed region.  
The same material body is compared in the two states; the transformed nucleus differs from the matrix by a small transformation strain and, in general, by small changes in elastic moduli.  
To first order in the transformation strain, the elastic contribution is local and depends only on the untransformed equilibrium fields.
We show in Section~\ref{sec:theory} that the leading-order contribution contains a term governed by the contraction between the pre-existing stress and deformation gradient, together with additional stored-energy terms arising from stiffness contrast.

The framework presented in this work is not restricted to classical nucleation theory.  
It gives the elastic contribution to the formation energy of a transformed region in a finitely deformed soft solid, and therefore applies whenever a local transformation produces a transformation strain or elastic contrast.  
This includes both non-conserved order-parameter transformations, such as crystallization or nematic ordering \cite{warner2007liquid, fried2005orientational}, and broader transformed-region problems involving phase separation, growth, or curing, where transport or reaction kinetics may also enter \cite{suslick2023frontal, li2024mechanical, LI2026106681, li2022nonlinear, matsuo1992patterns, zhou2023mechanics}.

The rest of the paper is organized as follows. 
 Section~\ref{sec:theory} sets up the finite-deformation energy difference and the effects of stress on CNT. 
 Section~\ref{sec:examples} specializes the results to a  compressible neo-Hookean material and presents the results for various loadings.  Section~\ref{sec:discussion} presents quantitative predictions for stress effects on CNT.  
 We conclude with Section~\ref{sec:conclusions}.
Detailed algebra is deferred to the appendices wherever appropriate.

\section{Elastic energy change due to nucleus formation}
\label{sec:theory}

\subsection{Kinematics}
Let $\Omega_{\rm R}$ denote the stress-free reference configuration of the soft solid.
This configuration is subjected to temperature change, $\Delta T$, and external tractions, $\bft^{0}$, that cause it to deform.
We denote the deformed configuration by $\Omega$.
The material points $\bfX \in \RefDom$ map to $\bfx \in \CurDom$ via the deformation map $\chicur$:
\begin{equation}
	\bfx = \chicur(\X),
\end{equation}
where
\begin{equation}
	\chicur: \RefDom \rightarrow \CurDom \quad \text{and}\quad \CurDom =\chicur(\RefDom).
\end{equation}

The corresponding deformation gradient, $\bfF$, maps infinitesimal line segments in $\RefDom$ to their deformed counterparts in $\Omega$
\begin{equation}
	\F \coloneqq  \pards{\xcur}{\bfX} = \pards{\chicur(\X)}{\bfX}.
\end{equation}
This map comprises the thermal and phase transformation induced deformation, $\bfF^{\rm th}$, as well as the accompanying elastic accommodation, $\bfF^{\rm e}$, and can be expressed using the Kr\"oner--Lee decomposition
\begin{equation}
	\F =  \Fe\Fth.
\end{equation}

\noindent We consider isotropic thermal deformations of the following form
\begin{equation}\label{eq:Fth_eta}
	\Fth(\bfX)
=
\begin{cases}
a(\bfX)(1+\eta)\id, & \X\in\SeedRef,\\
a(\bfX)\id, & \X\in\RefDom\setminus\SeedRef,
\end{cases}
\end{equation}
where $a(\bfX)$ represents the isotropic, temperature-change dependent thermal expansion, and $\eta \ll 1$
represents the isotropic phase-change expansion in a small nucleus $\SeedRef \subset \RefDom$ that undergoes phase transformation.

\begin{figure}[h]
  \centering
  \includegraphics[width = \linewidth]{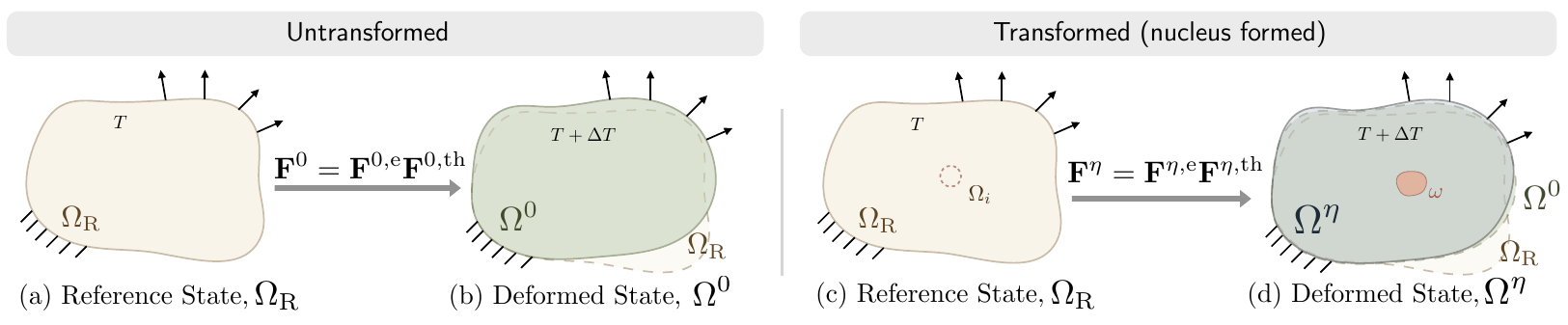}
  \caption{Kinematic states used in the calculation of energy change when a region in the parent material undergoes phase transformation.  (a) Stress-free reference configuration $\Omega_{\rm R}$.  (b) Untransformed, thermoelastically deformed state, $\Omega^0$, resulting from temperature change and applied tractions.  (c) Reference configuration showing pre-image of the transformed region, $\Omega_i \subset \Omega_R$. (d) Final deformed state, $\Omega^{\eta}$, after elastic accommodation of the transformation strain.}
  \label{fig:kinematic_states}
\end{figure}

We introduce further notation to distinguish between two scenarios (see Fig. \ref{fig:kinematic_states}), which will be important for calculating energy changes in \S\ref{sec:elastic_energy_change_main}.
In the first scenario, the deformations are only due to thermal expansion and all quantities are denoted by superscript $(\cdot)^0$. 
In the second scenario, both thermal and phase change expansions contribute to deformation and all quantities bear the superscript $(\cdot)^\eta$.
The corresponding Kr\"oner-Lee decompositions are
\begin{align}
	\F^{0}=&\FpreE\FpreTh \qquad &\text{[thermal only]}\\
	\F^{\eta}=&\Fetael\Fetath \qquad &\text{[thermal + phase transformation]}
\end{align}
In the limit $\eta \rightarrow 0$, both scenarios converge, and so do the corresponding field i.e. $\F^{\eta} \rightarrow \F^{0}, \Fetael \rightarrow \FpreE$, and $\Fetath \rightarrow \FpreTh$.

\subsection{Elastic energy change}
\label{sec:elastic_energy_change_main}
Our aim is to obtain the elastic contribution to the driving force for nucleation.
This contribution comes from the change in equilibrium potential energy when a small material region of the soft solid, $\Omega_i$, undergoes phase transformation.
The transformed region differs from the parent material in two ways: it carries a local transformation strain $\eta \ll 1$, and its elastic moduli may also differ from those of the parent phase. 
We will consider the material properties of the transformed region to be parametrized by $\eta$.

The above physics is captured by choosing the strain energy density function $\psi(\F^{\rm e},\eta)$ to depend on two arguments:
while $\F^{\rm e}$ accounts for the dependence of strain energy on elastic deformation, the explicit dependence on $\eta$ accounts for the phase-dependence of material properties.
In equilibrated states of the solid, $\F^{\rm e}$ also depends on $\eta$.
Thus, the equilibrium strain energy depends on $\eta$ through both arguments.

To calculate the change in elastic potential energy, we will first obtain the elastic deformation gradients in transformed and untransformed states, followed by calculation of potential energies in both states, and ultimately, their difference.

For our chosen form of thermal-transformation strain (see equation \eqref{eq:Fth_eta}), the corresponding elastic deformation gradient is
\begin{equation}
    \Fetael=\F^\eta(\Fetath)^{-1} =
    \begin{cases}
    \displaystyle
    \frac{1}{a(\bfX)(1+\eta)}\,\F^\eta,
    & \bfX\in\SeedRef,\\[8pt]
    \displaystyle
    \frac{1}{a(\bfX)}\,\F^\eta,
    & \bfX\in\RefDom\setminus\SeedRef .
    \end{cases}
    \label{eq:Fe_eta_explicit}
\end{equation}

Setting $\eta=0$, gives the elastic deformation gradient in the untransformed state
\begin{equation}
    \FpreE=\Fpre(\FpreTh)^{-1} = 
    \frac{1}{a(\bfX)}\Fpre. \label{eq:Fe_0_explicit}
\end{equation}

\noindent Next, we define the \textit{equilibrium} elastic potential energy of a solid containing a transformed region and subjected to dead-load tractions $\bft^0$ as
\begin{equation}\label{Pi_eta}
	\Pi^\eta = \int_{\Omega_{\rm R}\setminus \Omega_{\rm i}   } \psi(\Fetael,0) dV + \int_{\Omega_i} \psi(\Fetael,\eta) dV - \int_{\partial\RefDom}
    \bft^0\cdot\bfu^\eta\,dA,
\end{equation}
where $\bfu^\eta$ is the equilibrium displacement field, and $\Fetael$ is the elastic part of the equilibrium deformation gradient $\F^\eta$.\\

\noindent Similarly, the equilibrium elastic potential energy of the untransformed system ($\eta=0$) is 
\begin{equation}\label{Pi_0}
	\Pi^0 = \int_{\Omega_{\rm R}} \psi(\FpreE,0) dV - \int_{\partial\RefDom}
    \bft^0\cdot\bfu^0\,dA,
\end{equation}
where $\bfu^0$ is the equilibrium displacement field, and $\FpreE$ is the elastic part of the equilibrium deformation gradient $\F^0$.\\

\noindent Finally, the change in the equilibrium elastic potential energies between the transformed and untransformed states is obtained from Eq. \eqref{Pi_eta} and Eq. \eqref{Pi_0} as (see \ref{app:A} for a detailed derivation)
\begin{equation}
	\Delta \Pi(\eta) = \Pi^\eta -\Pi^0 = -
	\eta\int_{\Omega_i}\bfP^{0}\tendot\Fz dV + \eta\int_{\Omega_i}\left.\pards{\psi(\Fetael,\eta)}{\eta}\right|_{\Fetael = \FpreE, \eta = 0} dV + \mathcal{O}(\eta^2),\label{energy_diff}
\end{equation}
where 
\begin{equation*}
	\bfP^{0} = \left.\pards{\psi(\Fetael,\eta)}{\Fetael}\right|_{\Fetael = \FpreE, \eta = 0} (\FpreTh)^{-\top}
\end{equation*} 
is the first Piola--Kirchhoff stress in the untransformed state and will be the stress measure used throughout this study\footnote{Equivalently, using \(\mathbf P^0:\mathbf F^0=J^0\operatorname{tr}\boldsymbol\sigma^0\), the first integral may be expressed over the corresponding untransformed deformed region as
\(-\eta\int_{\omega}\operatorname{tr}\boldsymbol\sigma^0\,dv\).}.
Equation \eqref{energy_diff} is the main result of our work.
The first order energy difference has two contributions: the first term is due to the transformation strain of the nucleus, and the second term is due to the contrast in elastic properties between the nucleus and the parent material.
Notably, the energy difference depends only on the fields in the untransformed state in the region $\Omega_i \subset \OmegaR$ and on the magnitude of the transformation strain.\\

\noindent \myblue{\textbf{Remark \S2.1} If the transformation strain inside the inclusion is anisotropic instead of isotropic, i.e. $\bfF^{\rm th} = a(\bfX)(\id + \eta \bfA) $, where the tensor $\bfA$ brings in the anisotropy in transformation strain, the result in equation \eqref{energy_diff} generalizes as $\Delta \Pi(\eta) = -
	\eta\int_{\Omega_i}\bfP^{0}\tendot(\Fz \bfA) dV + \eta\int_{\Omega_i}\left.\pards{\psi(\Fetael,\eta)}{\eta}\right|_{\Fetael = \FpreE, \eta = 0} dV + \mathcal{O}(\eta^2)$.}

\subsection{Stress-modified classical nucleation theory}
\label{sec:cnt}
\noindent We now have the ingredients necessary to include the effects of stress in the classical nucleation theory.
\noindent The change in free energy upon formation of a spherical phase-transformed nucleus of radius $R$ is
\begin{equation}
  \DG(R)=4\pi R^2\gamma- \frac{4\pi}{3}R^3\Delta f_{\rm eff}.
  \label{eq:CNT_modified_main}
\end{equation}
where $\gamma$ is the surface energy of the nucleus-matrix interface and $\Delta f_{\rm eff}$ is the bulk portion of the free energy change.
The sign convention is that \(\Delta f_{\rm eff}>0\) favors growth of the transformed phase.  
The bulk driving force comprises thermal and elastic driving forces
\begin{equation}
\Delta f_{\rm eff}
=\underbrace{L\frac{\Tm-T}{\Tm}}_{\text{Thermal}}
+\underbrace{\eta\,\left(\Pz\tendot\Fz
-\left.\pards{\psi(\Fetael,\eta)}{\eta}\right|_{\Fetael = \FpreE, \eta = 0}\right) + O(\eta^2)}_{\text{Elastic}}
\label{eq:Dfeff_general_main}
\end{equation}
where \(L\) is the latent heat per unit reference volume.
The last term in the elastic contribution appears only when the transformed nucleus has elastic moduli different from those of the parent phase.  

The critical radius and nucleation barrier retain the classical three-dimensional spherical-CNT form:
\begin{equation}
  R^*=\frac{2\gamma}{\Delta f_{\rm eff}},
  \qquad
  \Delta G^*=\frac{16\pi\gamma^3}{3\Delta f_{\rm eff}^2}.
\label{eq:CNT_radius_barrier_main}
\end{equation}
These expressions apply only when \(\Delta f_{\rm eff}>0\).  When \(\Delta f_{\rm eff}\le 0\), the transformed phase is not locally favored by the bulk driving force.

Finally, the stress-shifted melting condition follows from \(\Delta f_{\rm eff}=0\)
\begin{equation}
  T_m^*
  =T_m + \eta\frac{T_m}{L}\left[
  \Pz\tendot\Fz
-\left.\pards{\psi(\Fetael,\eta)}{\eta}\right|_{\Fetael = \FpreE, \eta = 0}\right] +O(\eta^2).
\label{eq:Tm_shift_NH_main}
\end{equation}

In the following sections we will illustrate the results from the general theory by specializing it to a neo-Hookean material model, and applying it to uniform loading cases.\\

\noindent \textbf{Remark \S2.2} The linear dependence of bulk thermal energy on undercooling in \eqref{eq:Dfeff_general_main} is used only to make the connection with CNT explicit; the elastic calculation is unchanged if $\Delta f$ is replaced by a more general bulk thermodynamic driving force.

\section{Specialization to a neo-Hookean material}
\label{sec:examples}
We will specialize Eq.~\eqref{energy_diff} to a compressible neo-Hookean solid that is subjected to uniform thermal stretch \myblue{($a(\bfX)$ is taken to be constant)} and remote hydrostatic traction.

The strain energy density, $\psi$, has the following decoupled volumetric--isochoric form
\begin{equation}
  \psi(\Fe,\eta)
  =
  \frac{\mu(\eta)}{2}
  \left[
    J_e^{-2/3}\Fe\tendot\Fe -3
  \right]
  +
  \frac{K(\eta)}{2}\bigl(\ln J_e\bigr)^2,
  \qquad
  J_e=\det\Fe .
  \label{eq:NH_energy_main}
\end{equation}

The moduli of the transformed material $\{\mu(\eta), K(\eta) \}$ are assumed to differ perturbatively from those of the untransformed material  $\{\mu_0, K_0 \}$,
\begin{equation}
  \mu(\eta)=\mu_0+\eta\mu_1
  \equiv \mu_0+\delta\mu,
  \qquad
  K(\eta)=K_0+\eta K_1
  \equiv K_0+\delta K.
  \label{eq:NH_moduli_main}
\end{equation}

The thermal deformation gradient is taken to be constant, $\FpreTh=a\id$.
In the untransformed state, the elastic deformation is then easily related to the total deformation gradient as $\FpreE=\frac{1}{a}\Fz$ from Eq. \eqref{eq:Fe_0_explicit}. Noting that  $J^0=\det\Fz$, the first Piola--Kirchhoff stress is

\begin{equation}
\bfP^{0}
=
\mu_0 (J^0)^{-2/3}
\left[
\F^0-\frac{\F^0\tendot\F^0}{3}(\F^0)^{-\top}
\right]
+
K_0\ln\left(\frac{J^0}{a^3}\right)(\F^0)^{-\top}.
\label{eq:P_NH_F0_main}
\end{equation}

\noindent For a locally uniform nucleus of reference volume $V_i$, Eq.~\eqref{energy_diff} gives the mixed stress--stretch form
\begin{equation}
\Delta\Pi
=
V_i\left[
-\eta\,\Pz\tendot\Fz
+\frac{\delta \mu}{2}\left\{(J^0)^{-2/3}\Fz\tendot\Fz-3\right\}
+\frac{\delta K}{2}
\left\{\ln\!\left(\frac{J^0}{a^3}\right)\right\}^2
\right] +O(\eta^2).
\label{eq:DeltaPi_NH_mixed_main}
\end{equation}
This mixed stress--stretch form is convenient under prescribed  stress, because $\Pz\tendot\Fz$ can be evaluated directly from the applied stress and the resulting stretch\footnote{For the same neo-Hookean specialization, Eq.~\eqref{eq:DeltaPi_NH_mixed_main} may also be written entirely in terms of stretches: $
\Delta\Pi
=V_i\left[
-3\eta K_0\ln\!\left(\frac{J^0}{a^3}\right)
+\frac{\delta \mu}{2}\left\{(J^0)^{-2/3}\Fz\tendot\Fz-3\right\}
+\frac{\delta K}{2}\left\{\ln\!\left(\frac{J^0}{a^3}\right)\right\}^2
\right]$.}.
It is worth noting that in the energy change, the transformation strain couples to \(\Pz\tendot\Fz\), whereas shear- and bulk-modulus contrasts couple with the stored distortional and volumetric energies of the untransformed state.\\

\noindent For remote tensile hydrostatic loading of magnitude $S$, the stress and deformation gradient fields are of the form
\begin{equation}
\F^0=\lambda \id , \quad 	\bfP^0=S \id,
\end{equation}
where the two fields are connected from equation \eqref{eq:P_NH_F0_main}
\begin{equation}
	S =  \frac{3K_0}{\lambda}  \ln\left(\frac{\lambda}{a}\right).\label{stress-stretch-hydro}
\end{equation}
The resulting potential energy change then becomes
\begin{equation}
\Delta \Pi =V_i\!\left[
-3\eta S\lambda
+\frac{\delta K}{2}\left\{3\ln\!\left(\frac{\lambda}{a}\right)\right\}^2
\right] + O(\eta^2) = V_i\left[-3\eta S\lambda
+\frac{S^2 \lambda^2\delta K}{2K_0^2}\right] + O(\eta^2)
\end{equation}
where the second equality follows from equation \eqref{stress-stretch-hydro}.
Using this result together with equations \eqref{eq:Dfeff_general_main} and \eqref{eq:Tm_shift_NH_main} gives the total bulk driving force and stress-shifted melting point as

\begin{equation}
\Delta f_{\rm eff}
=L\frac{\Tm-T}{\Tm}
+\eta\,\left(3 S\lambda
-\frac{S^2 \lambda^2K_1}{2K_0^2}\right) + O(\eta^2)
\label{eq:Dfeff_general_main_explicit}
\end{equation}
and
\begin{equation}
  T_m^*
  =T_m + \eta\frac{T_m}{L}\left[
  3 S\lambda
-\frac{S^2 \lambda^2K_1}{2K_0^2}\right] +O(\eta^2).
\label{eq:Tm_shift_NH_main_explicit}
\end{equation}\\

\noindent \textbf{Remark \S3.1} Equation~\eqref{energy_diff} does not \textit{require} the thermal deformation $a$ to be constant or the stress $\bfP^0$ or the deformation $\bfF^{0}$ to be locally uniform in the nucleus. It also does not require the transformed region to be small. These assumptions are employed only to facilitate analytical estimates.\\

\noindent \textbf{Remark \S3.2} Hydrostatic loading is a special case where the shear-modulus contrast term vanishes.
Thus, for hydrostatic loading, the first order approximation can be applied to freezing-type phase changes where the shear modulus contrast is large but bulk modulus contrast is not.\\

\noindent \textbf{Remark \S3.3} The asymptotic analysis is valid only if the expansion is well ordered i.e. the $O(\eta)$ term is much bigger than $O(\eta^2)$ term.
To mathematically define this requirement, the second order term can be explicitly calculated for the special case of hydrostatic loading.
This result is derived in \ref{app:hydro_accommodation}.\\

\noindent We also present briefly the energy expressions for two other loading types---uniaxial and equibiaxial---in Table \ref{tab:DeltaPi_PK_cases}.

\begin{table}[H]
\caption{First-order formation energy for a locally uniform transformed region of reference volume $V_i$ under prescribed first Piola--Kirchhoff stress.  The thermal stretch is uniform, $\FpreTh=a\id$.  The stretches $\lambda$ and $\lambda_\perp$ are determined from the prescribed traction boundary conditions in \ref{app:diagonal_states}.  The matched-stiffness terms reduce to $-\eta V_i\Pz\tendot\Fz$; stiffness contrast adds the stored distortional and volumetric energy terms shown in the third column.}\label{tab:DeltaPi_PK_cases}
\centering
\small
\setlength{\tabcolsep}{3pt}
\renewcommand{\arraystretch}{1.35}
\begin{tabular*}{\textwidth}{@{\extracolsep{\fill}}lll@{}}
\hline
Loading
&
Matched stiffness
&
Small stiffness contrast
\\
\hline
\makecell[l]{Uniax.\\
$\bfP^0=\diag(S,0,0),\quad \Fpre=\diag(\lambda,\lambda_\perp,\lambda_\perp)$}
&
$\displaystyle -\eta V_i S\lambda$
&
$\displaystyle V_i\!\left[
-\eta S\lambda
+\frac{\delta\mu}{2}\left\{
\frac{\lambda^2+2\lambda_\perp^2}{(\lambda\lambda_\perp^2)^{2/3}}-3
\right\}
+\frac{\delta K}{2}\left\{\ln\!\left(\frac{\lambda\lambda_\perp^2}{a^3}\right)\right\}^2
\right]$
\\[1.0em]
\makecell[l]{Equibiax.\\
$\bfP^0=\diag(S,S,0),\quad \Fpre=\diag(\lambda,\lambda,\lambda_\perp)$}
&
$\displaystyle -2\eta V_i S\lambda$
&
$\displaystyle V_i\!\left[
-2\eta S\lambda
+\frac{\delta \mu}{2}\left\{
\frac{2\lambda^2+\lambda_\perp^2}{(\lambda^2\lambda_\perp)^{2/3}}-3
\right\}
+\frac{\delta K}{2}\left\{\ln\!\left(\frac{\lambda^2\lambda_\perp}{a^3}\right)\right\}^2
\right]$
\\[1.0em]
\makecell[l]{Hydro.\\
$\bfP^0=S\id,\quad \Fpre=\lambda\id$}
&
$\displaystyle -3\eta V_i S\lambda$
&
$\displaystyle V_i\!\left[
-3\eta S\lambda
+\frac{\delta K}{2}\left\{3\ln\!\left(\frac{\lambda}{a}\right)\right\}^2
\right]$
\\
\hline
\end{tabular*}
\end{table}

\section{Results and Discussion}
\label{sec:discussion}
We now evaluate the stress-modified nucleation theory for representative soft-solid material parameters.  
Unless otherwise stated, the calculations use
\[
K_0=10\,\mathrm{MPa},\qquad
\mu_0=1\,\mathrm{MPa},\qquad
L=2\times 10^8\,\mathrm{J/m^3},\qquad
T_m=273.15\,\mathrm{K},
\]
with an undercooling \(T_m-T=1\,\mathrm{K}\).  
The transformation strain amplitudes are \(\eta=\{0.01,0.02,0.03\}\).  
The  stress \(S\) is taken positive in tension.  
Hydrostatic results are plotted against \(S/K_0\), while uniaxial and equibiaxial results are plotted against \(S/\mu_0\).  
Dashed curves denote the stiffness-matched case, \(\delta\mu=\delta K=0\), for which the first-order mechanical contribution comes only from the \(\Pz\tendot\Fz\) term.  
Solid curves include an \(O(\eta)\) stiffness contrast, with \(\delta\mu=\eta\mu_0\) and \(\delta K=\eta K_0\), which gives rise to additional terms in the energy change (see Table \ref{tab:DeltaPi_PK_cases}).
For hydrostatic loading, the distortional change is identically zero, so the shear-modulus contrast does not contribute at first order and only \(\delta K=\eta K_0\) affects the plotted solid curves.
All critical radii and barriers are normalized by their zero-stress, stiffness-matched reference values,
\[
R_c^0=\frac{2\gamma}{\Delta f},
\qquad
\Delta G_c^0=\frac{16\pi\gamma^3}{3\Delta f^2},
\]
where \(\Delta f=L(T_m-T)/T_m\).  
Because only normalized critical quantities are shown, the interfacial energy \(\gamma\) cancels from the plotted ratios.

\begin{figure}[H]
    \centering
    \includegraphics[width=\textwidth]{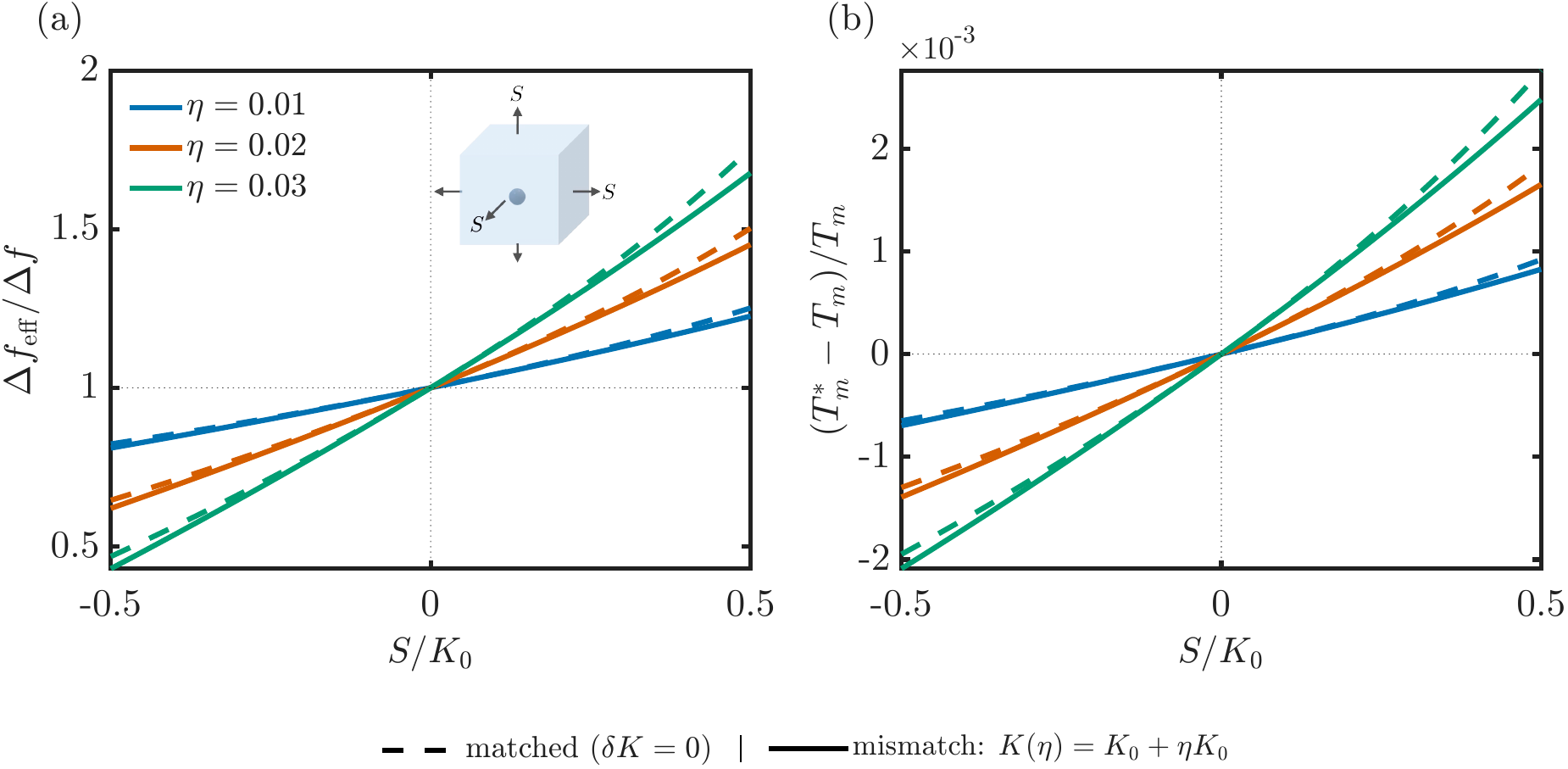}
    \caption{\textbf{Hydrostatic stress changes the effective driving force for phase transformation.}
(a) Effective driving force normalized by the zero-stress thermal driving force, (shown for undercooling of $1$\,K) and
(b) relative critical-temperature shift versus normalized  stress. Results are shown for $a=1$ and $\eta=\{0.01,0.02,0.03\}$.
Dashed curves denote matched stiffness; solid curves include $\delta K=\eta K_0$.
For hydrostatic loading, the shear-modulus contrast does not contribute at first order.}
    \label{fig:hydro_driving}
\end{figure}

Figure~\ref{fig:hydro_driving} shows the effect of hydrostatic stress on the effective bulk driving force and shifted transformation temperature for \(a=1\).  
Tensile stress increases the effective driving force for phase transformation, whereas compressive stress reduces it (for $\eta>0$).  
The response is not symmetric under \(S\mapsto -S\), because prescribed stress determines the hydrostatic stretch through
\[
S=\frac{3K_0}{\lambda}\ln\left(\frac{\lambda}{a}\right),
\]
and the stiffness-matched mechanical contribution scales as \(S\lambda\), not as \(S\) alone.  
The same energetic contribution shifts the transformation temperature upward in tension and downward in compression.  
The stiffness-contrast terms add positive stored-energy contributions to the formation energy and therefore partially offset the stress-induced increase in driving force.

\begin{figure}[H]
    \centering
    \includegraphics[width=\textwidth]{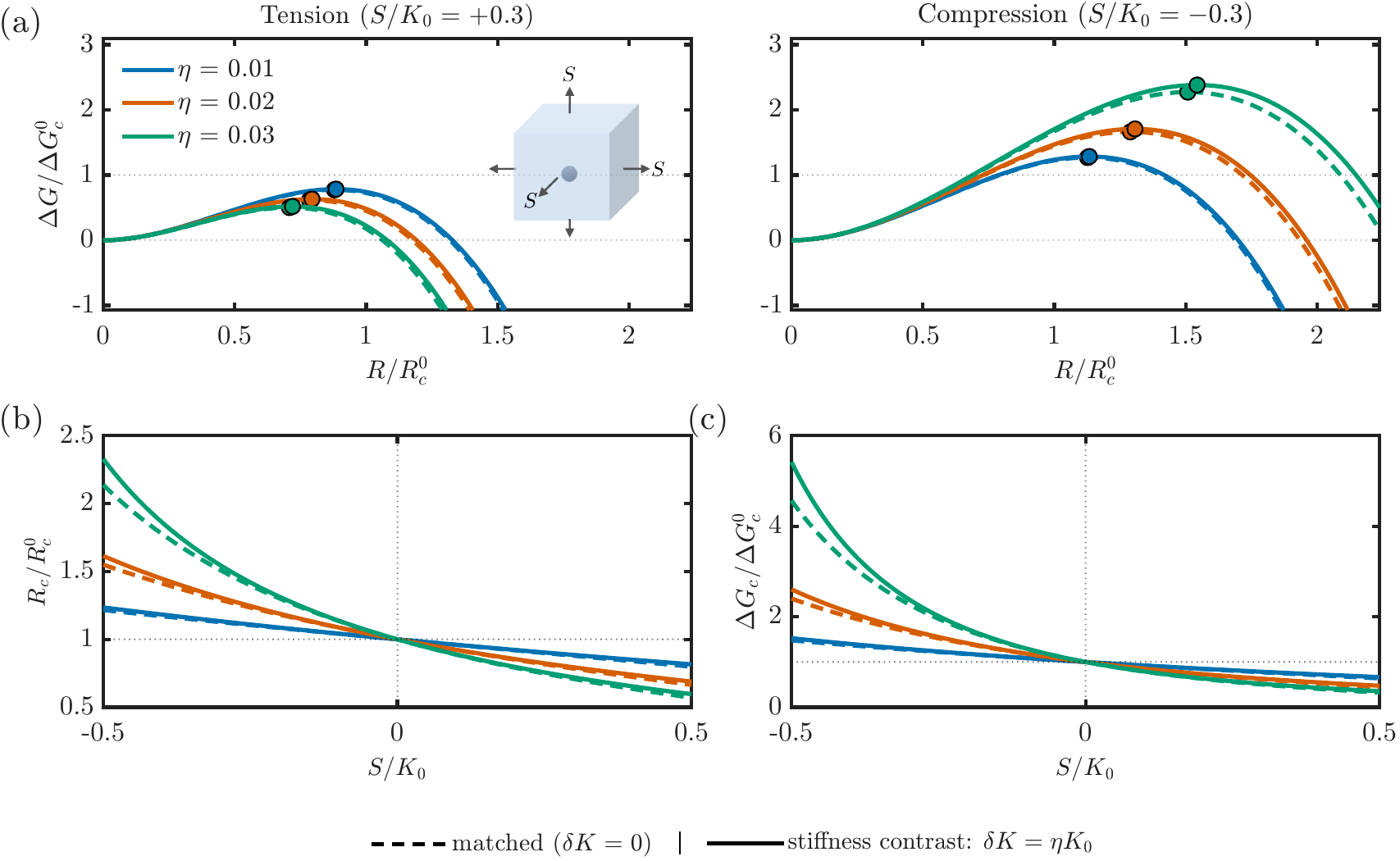}
\caption{\textbf{Hydrostatic stress changes the CNT critical radius and barrier.}
(a) Normalized free-energy landscapes at $S/K_0=\pm0.3$; markers indicate critical points.
(b) Critical-radius ratio and (c) nucleation-barrier ratio versus $S/K_0$.
Results use $a=1$, $T_m-T=1\,\mathrm{K}$, and $\eta=\{0.01,0.02,0.03\}$.
All quantities are normalized by their non-mechanical reference values, denoted by $(\cdot)^0$.
The reference critical radius and energy barrier are
$R_c^0=2\gamma/|\Delta f|$ and
$\Delta G_c^0=16\pi\gamma^3/(3|\Delta f|^2)$, respectively.
Dashed curves denote matched stiffness; solid curves include $\delta K=\eta K_0$.}
    \label{fig:hydro_landscape}
\end{figure}

The corresponding changes in the CNT free-energy landscape are shown in Fig.~\ref{fig:hydro_landscape}.  
For tensile hydrostatic loading, the increased driving force reduces both the critical radius and the nucleation barrier.  
For compressive loading, the driving force is reduced, so the critical radius and barrier increase.  
For hydrostatic loading, the stiffness-contrast contribution depends only on the bulk-modulus contrast, since the distortional stored-energy term vanishes on the hydrostatic branch.

\begin{figure}[H]
    \centering
    \includegraphics[width=\textwidth]{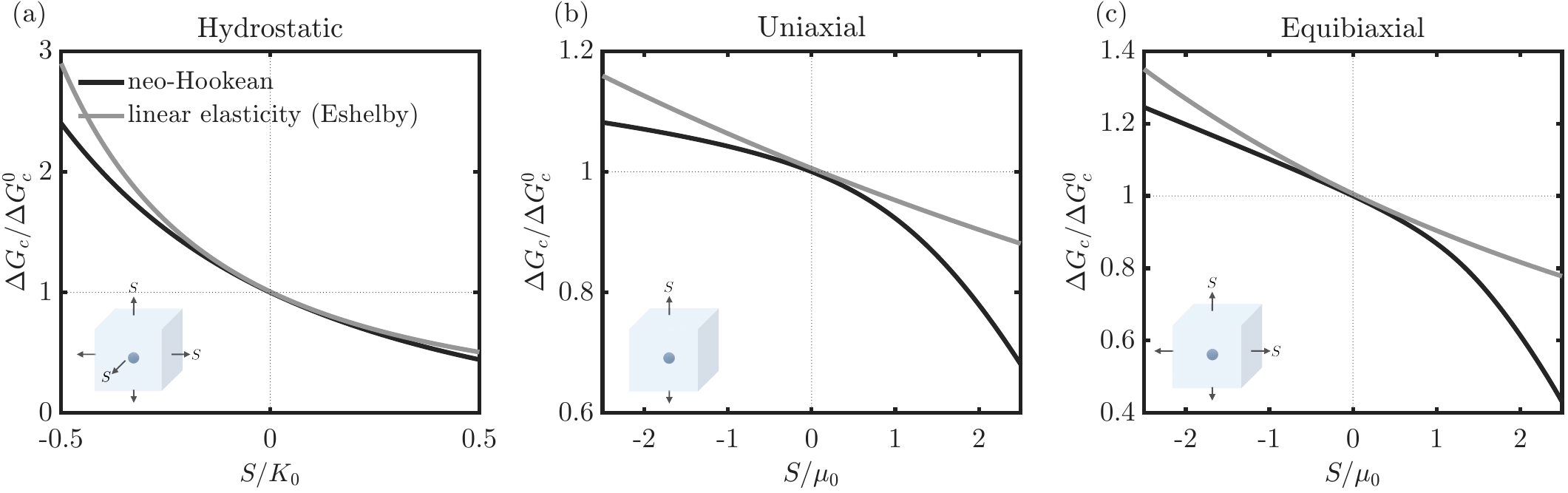}
\caption{\textbf{Linear elasticity overestimates nucleation barrier.}
Nucleation energy barrier ratio, $\Delta G_c/\Delta G_c^0,$ for $\eta=0.02$, $a=1$, and
$T_m-T=1\,\mathrm{K}$ under (a) hydrostatic, (b) uniaxial, and (c) equibiaxial  stress.
Black curves show the finite-deformation neo-Hookean prediction; gray curves show the
matched-stiffness linear-elastic Eshelby prediction.}
    \label{fig:FDvsLE}
\end{figure}

Figure~\ref{fig:FDvsLE} compares the finite-deformation neo-Hookean prediction with the matched-stiffness linear-elastic Eshelby theory prediction for \(\eta=0.02\).  
The comparison isolates the effect of finite deformation on the normalized barrier ratio \(\Delta G_c/\Delta G_c^0\) under hydrostatic, uniaxial, and equibiaxial loading.  
Linear elasticity, when applied beyond its validity at large stresses, overestimates the nucleation barrier as compared to the finite-deformation theory; this difference becomes more pronounced with increasing magnitude of stress.

Additional results are presented in \ref{App:D} that include uniaxial and equibiaxial loading as well as results for $a = 1.002$.
  Uniaxial and equibiaxial stresses can also modify the driving force, critical radius, and energy barrier considerably.
When present, the shear-modulus contrast contributes to the energy barrier for uniaxial and biaxial cases.
The results for \(a=1.002\) show that the same mechanisms persist when the untransformed body carries a finite thermal stretch.

Finally, in continuation of Remark \S3.3, a rigorous ordering estimate in \ref{app:hydro_accommodation} quantifies the validity of the first-order approximation for a spherical nucleus under remote hydrostatic loading.
For the parameters used here and \(\eta=0.02\), the ratio of $O(\eta^2)$ term and $O(\eta)$ term in the energy change remains small over a broad range of stress away from points where the first-order term vanishes or the incremental hydrostatic accommodation coefficient becomes singular.  
Thus the first-order theory is broadly applicable and serves as a useful analytical estimate; the appendix identifies regimes where second-order accommodation energy must be retained.

\section{\myblue{Direct numerical validation}}
\label{sec:FEvalidation}
{\color{myblue}
Predictions of the first-order estimate are validated by directly solving the nonlinear elasticity problem, beginning with the hydrostatic loading that is the focus of this work.
These calculations use the same parameters as used in the previous section: $K_0=10\,$MPa, $\mu_0=1\,$MPa, and $a=1$.
For hydrostatic remote stress the transformed-inclusion problem is spherically symmetric, so equilibrium reduces to a one-dimensional radial equation, which is numerically solved by an energy-minimizing radial discretization in MATLAB using a radially graded one-dimensional mesh.
The first-order energy of Table~\ref{tab:DeltaPi_PK_cases} matches this direct solution to a few percent for both the matched and the stiffness-contrast cases, as shown in Fig. 5(a)-(b) for $\eta =0.02$ case.
This confirms that the first-order formula is accurate to within a few percent except in a narrow window about $S=0$, where the leading term $-3\eta S\lambda$ vanishes and the second-order accommodation must be retained, consistent with the ordering requirement of Remark~\S3.3.
Further, for a representative case of $S/K_0 = 0.3$, we show in Fig. 5(c) that the absolute error between the asymptotic estimate and the numerically computed energy scales as $O(\eta^2)$, confirming the validity of the theory.
Within the small-$\eta$ range, therefore, the elastic part of the nucleation barrier follows from post-processing the untransformed state rather than from solving a new boundary-value problem for each nucleus.

Qualitatively similar conclusions hold for uniaxial and equibiaxial loading, computed with 3-D finite element simulations using \texttt{FEniCS} \cite{AlnaesEtal2015, LoggEtal2012, LoggWells2010, LoggEtal_10_2012, KirbyLogg2006, LoggEtal_11_2012, OlgaardWells2010, Kirby2004, kirby2010} package on a cubic matrix containing the spherical nucleus (see \ref{app:FE}, Figs.~\ref{fig:FEuni}--\ref{fig:FEbi} for details). 
}

\begin{figure}[H]
\centering
\includegraphics[width=\textwidth]{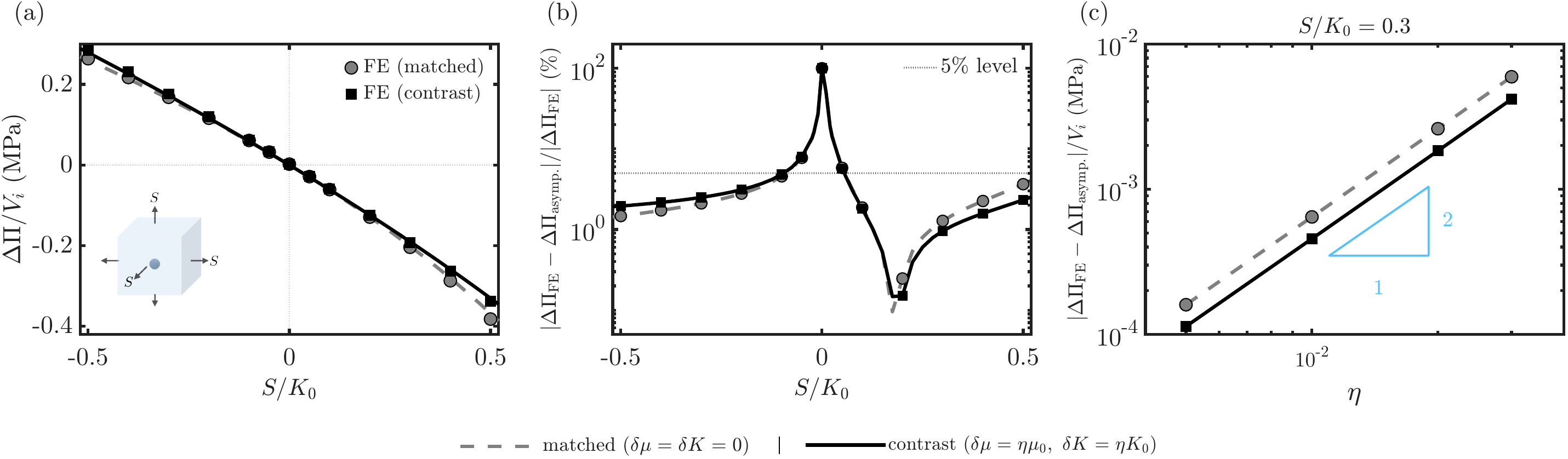}
\caption{\myblue{\textbf{Hydrostatic validation for matched and stiffness-contrast cases.} (a) $\Delta\Pi/V_i$ versus $S/K_0$ at $\eta=0.02$: first-order asymptotic (lines, Table~\ref{tab:DeltaPi_PK_cases}) and direct radial solution (markers). (b) Relative error of the first-order prediction with respect to the finite-element value; the dashed line marks the $5\%$ level (the peak at $S=0$ is where $\Delta\Pi_{\rm asymp.}\to0$, not a loss of accuracy---the absolute error there is the small $O(\eta^2)$ accommodation). (c) Absolute error $|\Delta\Pi_{\rm FE}-\Delta\Pi_{\rm asymp.}|/V_i$ versus $\eta$ at fixed stress $S/K_0=0.3$: the first-order asymptote converges to the direct solution as $O(\eta^2)$ (slope-2 reference triangle). Dashed gray curves show matched-stiffness and solid black curves show stiffness-contrast cases.}}
\label{fig:FEhydro}
\end{figure}

\section{Conclusions}
\label{sec:conclusions}
The central result is the first-order equilibrium elastic potential-energy change associated with forming a transformed region in a prestressed hyperelastic body:
\begin{equation}
\Delta \Pi
=
-\eta\int_{\Omega_i}\mathbf P^0:\mathbf F^0\,dV
+
\eta\int_{\Omega_i}
\left.
\frac{\partial \psi(\mathbf F^e,\eta)}{\partial \eta}
\right|_{\mathbf F^e=\mathbf F^{0,e},\,\eta=0}
dV
+O(\eta^2).
\end{equation}
Here \(\Omega_i\) is the reference-domain region occupied by the candidate transformed nucleus, which experiences an isotropic transformation strain $\eta\ll1$. 
The first integral is the work-conjugate contribution of the transformation strain in the pre-existing equilibrium state, while the second integral accounts for explicit phase-property dependence, including stiffness contrast between the parent and transformed phases.  
To first order in the transformation-strain amplitude, this energy change can be evaluated directly from the untransformed equilibrium stress and deformation fields.  Thus, the leading-order mechanical contribution does not require solving a new nonlinear elasticity boundary-value problem for each candidate nucleus.  
This is the main practical consequence of the theory: it converts the elastic part of the nucleation barrier into a post-processing calculation on a known prestressed state, within the range of validity of the small-transformation-strain expansion.

When incorporated into classical nucleation theory, this elastic energy change modifies the effective bulk driving force, transformation temperature, critical radius, and nucleation barrier.  
For the representative compressible neo-Hookean solid considered here with $\eta>0$, tensile prestress increases the driving force for the transformed phase and lowers the critical radius and barrier, whereas compressive prestress has the opposite effect.  
The response is not symmetric in tension and compression because the  stress, stretch, and elastic volume change are related nonlinearly at finite deformation.  
Stiffness contrast contributes additional stored-energy terms at the same asymptotic order when the changes in modului scale with the transformation-strain amplitude.

Comparison with the corresponding linear-elastic Eshelby-theory prediction shows that finite-deformation effects can substantially change the predicted energy barrier for phase transformation at moderate stretches.  
The difference arises because the finite-deformation theory retains the nonlinear relation between stress, deformation, and mechanical energy change, rather than replacing the prestressed state by its infinitesimal-strain approximation.

\myblue{It is worth delineating the regime of applicability of this result. 
Since the energy difference is calculated asymptotically, the first order estimate increases in accuracy as $\eta \rightarrow 0$. In practice, the theory is applicable to small transformation strains. 
The leading order energy difference vanishes when the prestress $S=0$; the dominant term is then $O(\eta^2)$, and the first-order estimate is correspondingly least accurate near $S=0$. 
For hydrostatic loading, the incremental elasticity problem loses strong ellipticity when $3K_0+4\mu_0-3\lambda S=0$. As this threshold is approached from the strongly elliptic regime, the second-order coefficient becomes singular and the asymptotic expansion ceases to be uniformly valid. For $3K_0+4\mu_0-3\lambda S<0$, the homogeneous prestressed state is no longer strongly elliptic, and the present nucleation analysis about that state is not physically meaningful (see \ref{app:hydro_accommodation}).}

Within this range of validity, the energy-difference calculation is broadly applicable: while the current study focuses on a non-conserved order parameter system, the calculation can be extended to conserved order parameter systems as well; the result provides a convenient and computationally efficient finite-deformation basis for incorporating stress-dependent nucleation barriers into simulations of phase-transforming soft solids.

The kinetic consequence is immediate;  once \(\Delta G_c\) is replaced by Eq.~\eqref{eq:CNT_radius_barrier_main}, any Arrhenius nucleation rate has the usual form \(J_{\rm nuc}\propto\exp[-\Delta G_c/(k_BT)]\).  
A full Avrami treatment that builds in the nucleation framework and tracks the long-time spatiotemporal evolution of a phase-transforming solid is beyond the scope of the current work and is deferred to a future study.
\myblue{The work can similarly be expanded to treat non-spherical nuclei as well as anisotropic interfacial energies to broaden its applicability to other material systems.}

\section*{Acknowledgments}
M.K. acknowledges support from the National Science Foundation CAREER Award (CMMI-2543637) and DEVCOM Army Research Laboratory (Award Number: W911NF-26-1-A041).

\newpage
\appendix

\section{Detailed derivation of the first order elastic energy change}\label{app:A}
To calculate the differences in the elastic potential energy between the transformed and untransformed states, we first obtain the changes in the elastic deformation gradient.
We perturb the general Kr\"oner-Lee decomposition, $\bfF^\eta = \Fetael\Fetath $ about the prestressed state using the following expansions:
\begin{equation}
	\bfF^\eta=\Fz+\delta\bfF+\mathcal{O}(\eta^2),
    \qquad
    \Fetael=\Fez+\delta\Fe+\mathcal{O}(\eta^2),
    \qquad
    \Fetath=\FpreTh+\delta\Fth+\mathcal{O}(\eta^2).
\end{equation}
where 
\begin{equation}
	\delta\bfF=O(\eta),
    \qquad
    \delta\Fe=O(\eta),
    \qquad
    \delta\Fth=O(\eta).
\end{equation}
Substituting these expansions in   $\bfF^\eta = \Fetael\Fetath$, we get  
\begin{equation}
    \delta\bfF
    =
    \delta\Fe\,\FpreTh
    +
    \Fez\,\delta\Fth,
\end{equation}
which can be rearranged to obtain
\begin{equation}
	\delta \Fe = (\delta\bfF -\Fez \delta\Fth)(\FpreTh)^{-1}.
\end{equation}
Together with equation \eqref{eq:Fth_eta} and noting that the transformation strain exists only inside the transformed nucleus, the elastic perturbation has the following piecewise form:
\begin{equation}\label{eq:dFe}
  \delta\Fe =
  \begin{cases}
    \frac{\delta\bfF - \eta\Fz}{\aX}, & \X\in \SeedRef\\
    \delta\bfF/\aX, & \X\in \Omega_{\rm m}.\\[4pt]
  \end{cases}
\end{equation}
\\
\noindent The elastic potential energies for the transformed and untransformed systems are
\begin{align}
\Pi^\eta =& \int_{\Omega_{\rm R}\setminus \Omega_{\rm i}} \psi(\Fetael,0) dV + \int_{\Omega_i} \psi(\Fetael,\eta) dV - \int_{\partial\Omega_{\rm R}}
    \bft^0\cdot\bfu^\eta\,dA, \qquad &\text{[transformed]}\\
	\Pi^0 =& \int_{\Omega_{\rm R}} \psi(\FpreE,0) dV - \int_{\partial\Omega_{\rm R}}
    \bft^0\cdot\bfu^0\,dA, \qquad &\text{[untransformed]}
\end{align} 
respectively.
Here, $\bfu^\eta$ and $\bfu^0$ are the equilibrium displacement fields, and $\Fetael$ and $\FpreE$ are elastic deformation gradients for transformed and untransformed systems, respectively.

To evaluate the difference in the potential energies, we consider first the difference in the strain energies, which we denote by $\Delta \psi$ and define as follows,
\begin{equation}\label{A8}
	\Delta \psi = \int_{\Omega_{\rm R}\setminus \Omega_{\rm i}} \psi(\Fetael,0) dV + \int_{\Omega_i} \psi(\Fetael,\eta) dV -
	 \int_{\Omega_{\rm R}} \psi(\FpreE,0) dV.
\end{equation}
Using equation \eqref{eq:dFe}, we get
\begin{multline}\label{strain_energy_diff1_appA_prelim}
	\Delta \psi =
	\int_{\Omega_{\rm R}\setminus \Omega_{\rm i}}\left.\pards{\psi(\Fetael,0)}{\Fetael}\right|_{\Fetael = \FpreE,\eta = 0}\tendot\frac{\delta\bfF}{a(\bfX)} dV + 
	\int_{\Omega_i}\left.\pards{\psi(\Fetael,\eta)}{\Fetael}\right|_{\Fetael = \FpreE,\eta = 0}\tendot\left(\frac{\delta\bfF- \eta\Fz}{\aX} \right) dV  \\ +
	 \eta\int_{\Omega_i}\left.\pards{\psi(\Fetael,\eta)}{\eta}\right|_{\Fetael = \FpreE, \eta = 0} dV +\mathcal{O}(\eta^2).
\end{multline}\\
Noting further that
\begin{equation}
	\bfP^{0,e} = \left.\pards{\psi(\Fetael,0)}{\Fetael}\right|_{\Fetael = \FpreE, \eta = 0}, \qquad \bfP^{0}= \bfP^{0,e}/a(\bfX),
\end{equation}
the difference in strain energy simplifies to
\begin{equation}\label{strain_energy_diff1_appA}
\Delta \psi =
	\int_{\Omega_{\rm R}}\bfP^{0}\tendot\delta\bfF dV -
	\eta\int_{\Omega_i}\bfP^{0}\tendot\Fz dV + \eta\int_{\Omega_i}\left.\pards{\psi(\Fetael,\eta)}{\eta}\right|_{\Fetael = \FpreE, \eta = 0} dV + \mathcal{O}(\eta^2).
\end{equation}

\noindent By application of divergence theorem on the first integral and invoking equilibrium, we find that 
\begin{equation}
	\int_{\Omega_{\rm R}}\bfP^{0}\tendot\delta\bfF dV = \int_{\partial\Omega_{\rm R}}\bft^{0}\cdot (\bfu^\eta - \bfu^0)\ dA,
\end{equation}
which nullifies the dead-load traction terms in the potential energy.\\

\noindent Thus, the potential energy difference reduces to

\begin{equation*}\boxed{
	\Pi^\eta -\Pi^0 = -
	\eta\int_{\Omega_i}\bfP^{0}\tendot\Fz dV + \eta\int_{\Omega_i}\left.\pards{\psi(\Fetael,\eta)}{\eta}\right|_{\Fetael = \FpreE, \eta = 0} dV + \mathcal{O}(\eta^2)}
\end{equation*}

{\color{myblue} \noindent \textbf{Note on anisotropic transformation strain.} If the transformation strain is anisotropic

\begin{equation}\label{eq:Fth_eta_aniso}
	\Fth(\bfX)
=
\begin{cases}
a(\bfX)(\id+\eta\bfA), & \X\in\SeedRef,\\
a(\bfX)\id, & \X\in\RefDom\setminus\SeedRef,
\end{cases}
\end{equation}

where $\bfA$ encapsulates the anisotropy, we get $\delta\Fth = \eta a(\bfX)\bfA$, which leads to

\begin{equation}\label{eq:dFe_aniso}
  \delta\Fe =
  \begin{cases}
    \frac{\delta\bfF - \eta\Fz\bfA}{\aX}, & \X\in \SeedRef\\
    \delta\bfF/\aX, & \X\in \Omega_{\rm m}.\\[4pt]
  \end{cases}
\end{equation}
Then along the same line of derivation following equation \eqref{A8}, we obtain the potential energy difference as

\begin{equation*}\boxed{
	\Pi^\eta -\Pi^0 = -
	\eta\int_{\Omega_i}\bfP^{0}\tendot(\Fz\bfA) dV + \eta\int_{\Omega_i}\left.\pards{\psi(\Fetael,\eta)}{\eta}\right|_{\Fetael = \FpreE, \eta = 0} dV + \mathcal{O}(\eta^2)}
\end{equation*}
for anisotropic transformation strain.
}

\section{Stress-stretch equations}
\label{app:diagonal_states}

We derive the equations used to calculate the stretches for prescribed uniaxial, equibiaxial, and hydrostatic first Piola--Kirchhoff stress states.  These equations support the loading-state results summarized in Table~\ref{tab:DeltaPi_PK_cases}.  
All three loading states are described by a diagonal total deformation gradient

\begin{equation}
  \Fpre=\diag(\lambda_1,\lambda_2,\lambda_3),
\end{equation}
where the principal stretches include the uniform thermal stretch \(a\).  Substitution into Eq.~\eqref{eq:P_NH_F0_main} gives the principal  stresses directly in terms of the stretches:
\begin{align}
P_{11}^0
&=
\frac{\mu_0}{3\lambda_1}
(\lambda_1\lambda_2\lambda_3)^{-2/3}
\left(
2\lambda_1^2-\lambda_2^2-\lambda_3^2
\right)
+
\frac{K_0}{\lambda_1}
\ln\left(
\frac{\lambda_1\lambda_2\lambda_3}{a^3}
\right),
\label{eq:P11_stretches}
\\
P_{22}^0
&=
\frac{\mu_0}{3\lambda_2}
(\lambda_1\lambda_2\lambda_3)^{-2/3}
\left(
2\lambda_2^2-\lambda_1^2-\lambda_3^2
\right)
+
\frac{K_0}{\lambda_2}
\ln\left(
\frac{\lambda_1\lambda_2\lambda_3}{a^3}
\right),
\label{eq:P22_stretches}
\\
P_{33}^0
&=
\frac{\mu_0}{3\lambda_3}
(\lambda_1\lambda_2\lambda_3)^{-2/3}
\left(
2\lambda_3^2-\lambda_1^2-\lambda_2^2
\right)
+
\frac{K_0}{\lambda_3}
\ln\left(
\frac{\lambda_1\lambda_2\lambda_3}{a^3}
\right).
\label{eq:P33_stretches}
\end{align}
The off-diagonal stress components vanish.

\paragraph*{Hydrostatic}
For hydrostatic stress control,
\[
  \bfP^0=S\id,
  \qquad
  \Fpre=\lambda\id .
\]
The three stress conditions \(P_{11}^0=P_{22}^0=P_{33}^0=S\) reduce to
\begin{equation}
S
=
\frac{3K_0}{\lambda}
\ln\left(
\frac{\lambda}{a}
\right).
\label{eq:hydrostatic_P_raw}
\end{equation}
Equivalently, \(\lambda\) is determined for a prescribed hydrostatic  stress \(S\) by
\begin{equation}
  \ln\left(
  \frac{\lambda}{a}
  \right)
  =
  \frac{S\lambda}{3K_0}.
  \label{eq:hydrostatic_log_equation}
\end{equation}

\paragraph*{Uniaxial}
For uniaxial stress control,
\[
  \bfP^0=\diag(S,0,0),
  \qquad
  \Fpre=\diag(\lambda,\lambda_\perp,\lambda_\perp).
\]
The stress conditions \(P_{11}^0=S\) and \(P_{22}^0=P_{33}^0=0\) give
\begin{align}
S
&=
\frac{2\mu_0}{3\lambda}
(\lambda\lambda_\perp^2)^{-2/3}
\left(
\lambda^2-\lambda_\perp^2
\right)
+
\frac{K_0}{\lambda}
\ln\left(
\frac{\lambda\lambda_\perp^2}{a^3}
\right),
\label{eq:uniaxial_P11_raw}
\\
0
&=
\frac{\mu_0}{3\lambda_\perp}
(\lambda\lambda_\perp^2)^{-2/3}
\left(
\lambda_\perp^2-\lambda^2
\right)
+
\frac{K_0}{\lambda_\perp}
\ln\left(
\frac{\lambda\lambda_\perp^2}{a^3}
\right).
\label{eq:uniaxial_P22_raw}
\end{align}
Equivalently, the following two equations determine \(\lambda\) and \(\lambda_\perp\) for a prescribed uniaxial stress \(S\):
\begin{equation}
S
=
\frac{\mu_0}{\lambda}
\left[
a^3
\exp\left(\frac{S\lambda}{3K_0}\right)
\right]^{-2/3}
\left[
\lambda^2
-
\frac{a^3}{\lambda}
\exp\left(\frac{S\lambda}{3K_0}\right)
\right],
\label{eq:uniaxial_single_equation}
\end{equation}
\begin{equation}
  \lambda_\perp
  =
  \left[
  \frac{a^3}{\lambda}
  \exp\left(\frac{S\lambda}{3K_0}\right)
  \right]^{1/2}.
  \label{eq:uniaxial_lambdaperp_eliminated}
\end{equation}

\paragraph*{Equibiaxial}
For equibiaxial stress control,
\[
  \bfP^0=\diag(S,S,0),
  \qquad
  \Fpre=\diag(\lambda,\lambda,\lambda_\perp).
\]
The stress conditions \(P_{11}^0=P_{22}^0=S\) and \(P_{33}^0=0\) give
\begin{align}
S
&=
\frac{\mu_0}{3\lambda}
(\lambda^2\lambda_\perp)^{-2/3}
\left(
\lambda^2-\lambda_\perp^2
\right)
+
\frac{K_0}{\lambda}
\ln\left(
\frac{\lambda^2\lambda_\perp}{a^3}
\right),
\label{eq:equibiaxial_P11_raw}
\\
0
&=
\frac{2\mu_0}{3\lambda_\perp}
(\lambda^2\lambda_\perp)^{-2/3}
\left(
\lambda_\perp^2-\lambda^2
\right)
+
\frac{K_0}{\lambda_\perp}
\ln\left(
\frac{\lambda^2\lambda_\perp}{a^3}
\right).
\label{eq:equibiaxial_P33_raw}
\end{align}
Equivalently, the following two equations determine \(\lambda\) and \(\lambda_\perp\) for a prescribed equibiaxial stress \(S\):
\begin{equation}
S
=
\frac{\mu_0}{\lambda}
\left[
a^3
\exp\left(\frac{2S\lambda}{3K_0}\right)
\right]^{-2/3}
\left[
\lambda^2
-
\frac{a^6}{\lambda^4}
\exp\left(\frac{4S\lambda}{3K_0}\right)
\right],
\label{eq:equibiaxial_single_equation}
\end{equation}
\begin{equation}
  \lambda_\perp
  =
  \frac{a^3}{\lambda^2}
  \exp\left(\frac{2S\lambda}{3K_0}\right).
  \label{eq:equibiaxial_lambdaperp_eliminated}
\end{equation}

\section{Hydrostatic spherical-nucleus accommodation at second order}
\label{app:hydro_accommodation}

This appendix estimates the second-order term in the formation energy for the hydrostatic case.  
The purpose is to quantify when the use of first-order result used in the main text is valid i.e. when is the asymptotic expansion of the energy difference well ordered.  
We consider a spherical transformed region of reference radius \(A\) embedded in a much larger spherical body, subjected to remote hydrostatic loading of magnitude $S$ at the remote boundary. The sign convention is tension-positive.\\

\noindent \textbf{Untransformed State.} In the untransformed state, the stress and deformation gradient in the body is homogeneous.

\begin{equation}
\F^0=\lambda \id , \quad 	\bfP^0=S \id.
\end{equation}
  
For reference, the first Piola--Kirchhoff stress is
\begin{equation}
\bfP^{0}
=
\mu_0 (J^0)^{-2/3}
\left[
\F^0-\frac{\F^0\tendot\F^0}{3}(\F^0)^{-\top}
\right]
+
K_0\ln\left(\frac{J^0}{a^3}\right)(\F^0)^{-\top},
\label{eq:P_NH_F0_app}
\end{equation}
where \(J^0=\det\F^0\).
Therefore, the stretch and applied loading are connected as
\begin{equation}
  S
  =
  \frac{K_0}{\lambda}
  \ln\left(\frac{\lambda^3}{a^3}\right).
  \label{eq:hydro_base_appendix}
\end{equation}
Equivalently,
\begin{equation}
  K_0 H=\lambda S,
  \qquad
  H\coloneqq \ln\left(\frac{\lambda^3}{a^3}\right).
  \label{eq:H_def_appendix}
\end{equation}
Here \(H\) is the logarithmic elastic volume change in the hydrostatic pre-state.\\

\noindent \textbf{Transformed State.}
The transformed nucleus is assigned
\[
  \F_i^{\rm th}=a(1+\eta)\id,
\]
whereas the matrix has
\[
  \F_m^{\rm th}=a\id .
\]
The nucleus moduli are expanded as
\begin{equation}
  K_i=K_0+\eta K_1,
  \qquad
  \mu_i=\mu_0+\eta \mu_1.
  \label{eq:hydro_moduli_appendix}
\end{equation}
Thus the actual modulus jumps are \(O(\eta)\): \(\delta K=\eta K_1\) and \(\delta\mu=\eta\mu_1\).\\

After the transformation, the radial deformation in the nucleus and matrix have different forms.
Spherical symmetry of the system ensures that the stress in the nucleus remains hydrostatic, which leads to a constant stretch.
Thus, we can take the radial deformation as
\begin{equation}
  r_i(R)=\lambda_i R\,
  \qquad 0\le R\le A.
  \label{eq:ri_appendix}
\end{equation}
Radial deformation in the matrix can be written as a perturbation of the hydrostatic pre-state,
\begin{equation}\label{eq:rm_asym}
r_m(R)=\lambda R(1+\eta u_m(R))+O(\eta^2).	
\end{equation}
 
By continuity of displacements at $R=A$, we get
\begin{equation}
	\lambda_i A = \lambda A(1 + \eta u_m(A))+O(\eta^2),
\end{equation}
which for convenience of representation can be restated as
\begin{equation}
	\lambda_i = \lambda(1 + \eta c)+O(\eta^2),
\end{equation}
where $c :=u(A)$.
Then the transformed interface is at 
\begin{equation}
 r_i(A) = r_m(A)=\lambda A(1+\eta c)+O(\eta^2),
  \label{eq:interface_c_appendix}
\end{equation}
where \(c\) is the first-order accommodation coefficient.  

The nucleus elastic stretch is therefore
\begin{equation}
  \lambda_i^{\rm e}
  =
  \frac{\lambda(1+\eta c)}{a(1+\eta)}
  =
  \frac{\lambda}{a}\left[1+\eta(c-1)\right]+O(\eta^2).
  \label{eq:seed_elastic_stretch_appendix}
\end{equation}
Since this deformation is spherical, \(\bar I_{1,i}^{\rm e}=3\).  Hence the nucleus shear-modulus contrast does not enter the hydrostatic result through this order.\\

The perturbation function $u_m$ is determined by satisfaction of equilibrium equation.
From equation \eqref{eq:P_NH_F0_main} and using the stretches in the matrix derived from equation \eqref{eq:rm_asym}, we obtain the stresses $P_r$ and $P_\theta = P_\phi$ and solve the equilibrium equation
\begin{equation}
\frac{dP_r}{dR}+\frac{2}{R}(P_r-P_\theta)=0	
\end{equation}
about the hydrostatic pre-state.  

The equilibrium equation at $O(\eta)$ gives
\begin{equation}
  \left(K_0+\frac{4\mu_0}{3}-\lambda S\right)  \left[  Ru_m''+4u_m'  \right]=0.  \label{eq:hydro_incremental_ode_appendix}
\end{equation}
{\color{myblue} When the coefficient \(K_0+4\mu_0/3-\lambda S\) vanishes, the system loses strong ellipticity.
Away from the incremental singularity and in the strong-ellipticity regime (\(K_0+4\mu_0/3-\lambda S > 0\))}, the decaying solution in the infinite-matrix limit is
\begin{equation}
  u_m(R)=c\frac{A^3}{R^3},
  \qquad R\ge A.
  \label{eq:matrix_displacement_appendix}
\end{equation}
Thus
\begin{equation}
  r_m(R)=  \lambda R\left(1+  \eta c\frac{A^3}{R^3}\right)  +  O(\eta^2).
  \label{eq:rm_appendix}
\end{equation}
The corresponding matrix stretches are
\begin{equation}
  \lambda_{r,m}
  =
  \lambda\left[
  1-2\eta c\frac{A^3}{R^3}
  \right]+O(\eta^2),
  \qquad
  \lambda_{\theta,m}
  =
  \lambda\left[
  1+\eta c\frac{A^3}{R^3}
  \right]+O(\eta^2).
  \label{eq:matrix_stretches_appendix}
\end{equation}

Traction continuity at \(R=A\) determines \(c\).  Expanding the nucleus and matrix radial first Piola stresses to \(O(\eta)\) gives
\begin{equation}
  P_{r,i}(A)
  =
  S
  +
  \frac{\eta}{\lambda}
  \left[
  K_1 H
  +
  3K_0(c-1)
  -
  c\lambda S
  \right]
  +
  O(\eta^2),
  \label{eq:seed_stress_appendix}
\end{equation}
and
\begin{equation}
  P_{r,m}(A)
  =
  S
  +
  \frac{\eta c}{\lambda}
  \left(
  -4\mu_0+2\lambda S
  \right)
  +
  O(\eta^2).
  \label{eq:matrix_stress_appendix}
\end{equation}
Setting \(P_{r,i}(A)=P_{r,m}(A)\) yields
\begin{equation}
  c
  =
  \frac{3K_0-K_1 H}
       {3K_0+4\mu_0-3\lambda S}.
  \label{eq:c_hydro_appendix}
\end{equation}
We now know all the relevant fields up to the first order.\\

\noindent \textbf{Energy Difference.} We can now compute the change in potential energy. Let,
\[
  V_i=\frac{4\pi A^3}{3}
\]
be the reference volume of the transformed nucleus.  The full potential energy consists of the nucleus energy, the matrix energy, and the dead-load potential.  The second-order displacement field is not needed for the \(O(\eta^2)\) coefficient because it enters only through the first variation of the base potential energy, which vanishes by equilibrium of the hydrostatic pre-state.

The nucleus contribution follows from
\[
  \psi_i
  =
  \frac{K_i}{2}\left(\log J_i^{\rm e}\right)^2,
\]
with
\begin{equation}
  \log J_i^{\rm e}
  =
  H
  +
  3(c-1)\eta
  -
  \frac{3}{2}(c^2-1)\eta^2
  +
  O(\eta^3).
  \label{eq:logJ_seed_appendix}
\end{equation}
Using \(K_i=K_0+\eta K_1+O(\eta^2)\) and \(K_0H=\lambda S\), we obtain
\begin{align}
  \frac{\Delta\Pi_i}{V_i}
  =
  &
  \left[
  3\lambda S(c-1)
  +
  \frac{1}{2}K_1H^2
  \right]\eta
  \nonumber\\
  &
  +
  \left[
  \frac{9}{2}K_0(c-1)^2
  -
  \frac{3}{2}\lambda S(c^2-1)
  +
  3K_1H(c-1)
  \right]\eta^2
  +
  O(\eta^3).
  \label{eq:seed_energy_appendix}
\end{align}

The matrix contribution is spatially nonuniform.  From Eq.~\eqref{eq:matrix_stretches_appendix},
\begin{equation}
  \psi_m-\psi_0
  =
  (6\mu_0-3\lambda S)c^2\eta^2\frac{A^6}{R^6}
  +
  O(\eta^3).
  \label{eq:matrix_density_appendix}
\end{equation}
Therefore
\begin{align}
  \Delta\Pi_m
  &=
  4\pi
  \int_A^\infty
  R^2
  (6\mu_0-3\lambda S)c^2\eta^2\frac{A^6}{R^6}
  \,dR
  +
  O(\eta^3)
  \nonumber\\
  &=
  V_i(6\mu_0-3\lambda S)c^2\eta^2
  +
  O(\eta^3).
  \label{eq:matrix_energy_appendix}
\end{align}
Finally, the first-order change in the dead-load potential is
\begin{equation}
  \frac{\Delta\Pi_{\rm load}}{V_i}
  =
  -3\lambda S c\,\eta
  +
  O(\eta^2),
  \label{eq:load_energy_appendix}
\end{equation}
with the \(O(\eta^2)\) terms involving the second-order displacement canceling by stationarity of the base state.

Combining Eqs.~\eqref{eq:seed_energy_appendix}--\eqref{eq:load_energy_appendix} gives
\begin{equation}
  \frac{\Delta\Pi}{V_i}
  =
  A_1\eta
  +
  A_2(c)\eta^2
  +
  O(\eta^3),
  \label{eq:energy_c_appendix}
\end{equation}
where
\begin{equation}
  A_1
  =
  -3\lambda S+\frac{1}{2}K_1H^2,
  \label{eq:A1_appendix}
\end{equation}
and
\begin{align}
  A_2(c)
  =
  &
  \frac{9}{2}K_0(c-1)^2
  -
  \frac{3}{2}\lambda S(c^2-1)
  +
  3K_1H(c-1)
  \nonumber\\
  &
  +
  (6\mu_0-3\lambda S)c^2 .
  \label{eq:A2c_appendix}
\end{align}
Substituting Eq.~\eqref{eq:c_hydro_appendix} gives the final second-order coefficient
\begin{equation}
  A_2
  =
  \frac{9}{2}K_0
  +
  \frac{3}{2}\lambda S
  -
  3K_1H
  -
  \frac{3}{2}
  \frac{(3K_0-K_1H)^2}
       {3K_0+4\mu_0-3\lambda S}.
  \label{eq:A2_final_appendix}
\end{equation}
Thus
\begin{equation}
\boxed{
  \frac{\Delta\Pi}{V_i}
  =
  \left[
  -3\lambda S+\frac{1}{2}K_1H^2
  \right]\eta
  +
  \left[
  \frac{9}{2}K_0
  +
  \frac{3}{2}\lambda S
  -
  3K_1H
  -
  \frac{3}{2}
  \frac{(3K_0-K_1H)^2}
       {3K_0+4\mu_0-3\lambda S}
  \right]\eta^2
  +
  O(\eta^3),
}
  \label{eq:hydro_energy_final_appendix}
\end{equation}
where
\[
  H=\ln\left(\frac{\lambda^3}{a^3}\right),
  \qquad
  K_0H=\lambda S.
\]\\

\noindent \textbf{Well-orderedness of the Energy Asymptotic Expansion.} Equation~\eqref{eq:hydro_energy_final_appendix} also makes the ordering requirement explicit.  The first-order approximation used in the main text is well ordered when
\[
  |A_2\eta|\ll |A_1|.
\]
If \(A_1\) is small, the second-order accommodation energy becomes the leading mechanical contribution.
Some observations:

\noindent 1. For zero remote loading,
\[
  S=0,
  \qquad
  \lambda=a,
  \qquad
  H=0.
\]
Then \(A_1=0\),
\[
  c=\frac{3K_0}{3K_0+4\mu_0},
\]
and Eq.~\eqref{eq:hydro_energy_final_appendix} reduces to
\begin{equation}
  \frac{\Delta\Pi}{V_i}
  =
  \frac{18K_0\mu_0}{3K_0+4\mu_0}\eta^2
  +
  O(\eta^3).
  \label{eq:hydro_zeroS_appendix}
\end{equation}
Clearly, the first order energy change is not the dominant term in this scenario. \\

\noindent 2. In the nearly incompressible limit \(K_0\gg \mu_0\) with the applied load $S$ being finite, 
\begin{equation}
	\lambda \rightarrow a, \quad H \rightarrow 0.
\end{equation}
Assuming that $K_1/K_0$ is finite, 
\begin{equation}
	A_1 \rightarrow -3\lambda S, \quad A_2 \rightarrow 6\mu_0 -3\lambda S.
\end{equation}
Thus, the energy change becomes
\begin{equation}
	\frac{\Delta \Pi}{V_i} = -3\lambda S\eta + (6\mu_0-3\lambda S)\eta^2 + O(\eta^3)
\end{equation}
Thus, the asymptotic energy expansion is well ordered when,
\begin{equation}
	|\eta| \ll \left|\frac{\lambda S}{2\mu_0 -\lambda S}\right|.
\end{equation}

\begin{figure}[H]
    \centering
    \includegraphics[width=0.6\textwidth]{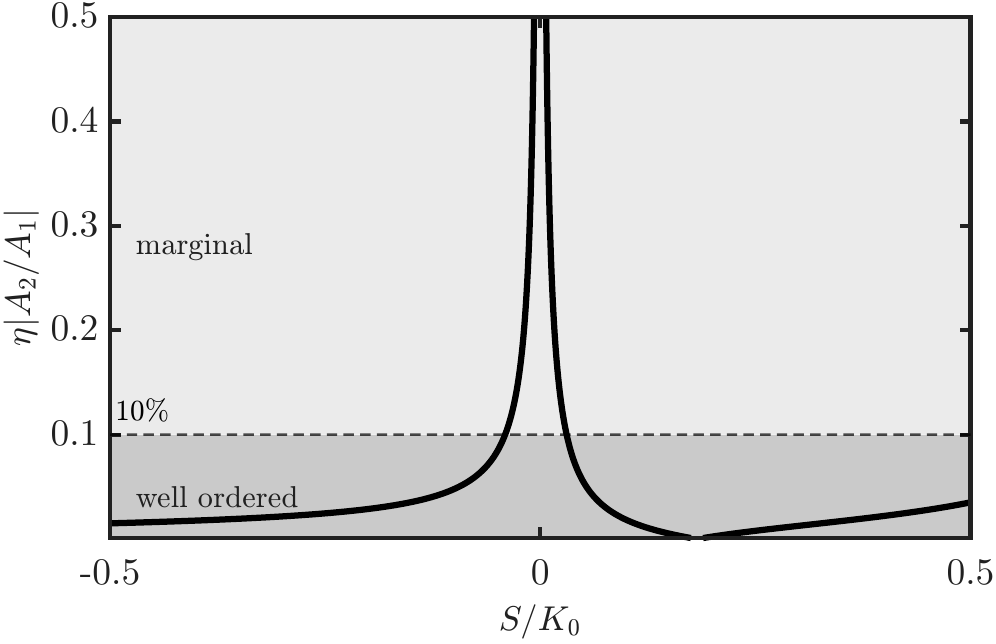}
    \caption{\textbf{Asymptotic Ordering of the Potential Energy Change for Hydrostatic Loading with Matched Stiffness.}
Ratio of the second and the first order term, $\eta |A_2/A_1|$,  is shown for $a=1$, $K_0=10$\,MPa,  $\mu_0=1$\,MPa, $K_1=0$, and $\mu_1=0$. The expansion is well ordered when $\eta |A_2/A_1| \ll 1$. The singularity at $S=0$ is because the first order term disappears. The first order approximation works well for a large range of pre-existing stress.
}\label{ordering}
\end{figure}

\noindent 3. For a matched-stiffness case, the results from equation \eqref{eq:hydro_energy_final_appendix} can be summarized in Table \ref{tab:hydro_log_ordering} and figure \ref{ordering}.
We find that for a wide range of stresses, the approximation is well ordered; the singularity that is seen at $S=0$ is already noted in the first observation.

\begin{table}[H]
\centering
\caption{Ordering of the hydrostatic expansion for the logarithmic volumetric law with $\eta=0.02$, $K_0=10\,\mathrm{MPa}$, $\mu_0=1\,\mathrm{MPa}$, $K_1=0$, $\mu_1=0$, and $a=1$. The expansion is well ordered when $\eta |A_2/A_1|\ll 1$. The singular point $S=0$ is omitted because $A_1=0$.}
\label{tab:hydro_log_ordering}
\begin{tabular}{c c c c}
\hline
$S/K_0$ & $\lambda$ & $|A_1/A_2|$ & $\eta |A_2/A_1|$ \\
\hline
$-0.50$ & $0.8656$ & $1.328$  & $0.01506$ \\
$-0.30$ & $0.9128$ & $0.9216$ & $0.02170$ \\
$-0.20$ & $0.9393$ & $0.6939$ & $0.02882$ \\
$-0.10$ & $0.9682$ & $0.4169$ & $0.04797$ \\
$\phantom{-}0.10$ & $1.0351$ & $1.087$  & $0.01839$ \\
$\phantom{-}0.30$ & $1.1183$ & $1.580$  & $0.01266$ \\
$\phantom{-}0.50$ & $1.2269$ & $0.5688$ & $0.03516$ \\
\hline
\end{tabular}
\end{table}

\section{Additional Results}
\label{App:D}
\noindent The material parameters are $L=2\times 10^8$\,J/m$^3$, $K_0=10$\,MPa, $\mu_0=1$\,MPa and $T_m=273.15$\,K.
For normalization, non-mechanical reference values, denoted by $(\cdot)^0$, are used throughout.
The reference critical radius and energy barrier are $R_c^0=2\gamma/|\Delta f|$ and
$\Delta G_c^0=16\pi\gamma^3/(3|\Delta f|^2)$, respectively.

\paragraph{\textbf{I. Uniaxial and Biaxial for a=1}} 
\subsection*{\textbf{Uniaxial}}
\begin{figure}[H]
    \centering
    \includegraphics[width=\textwidth]{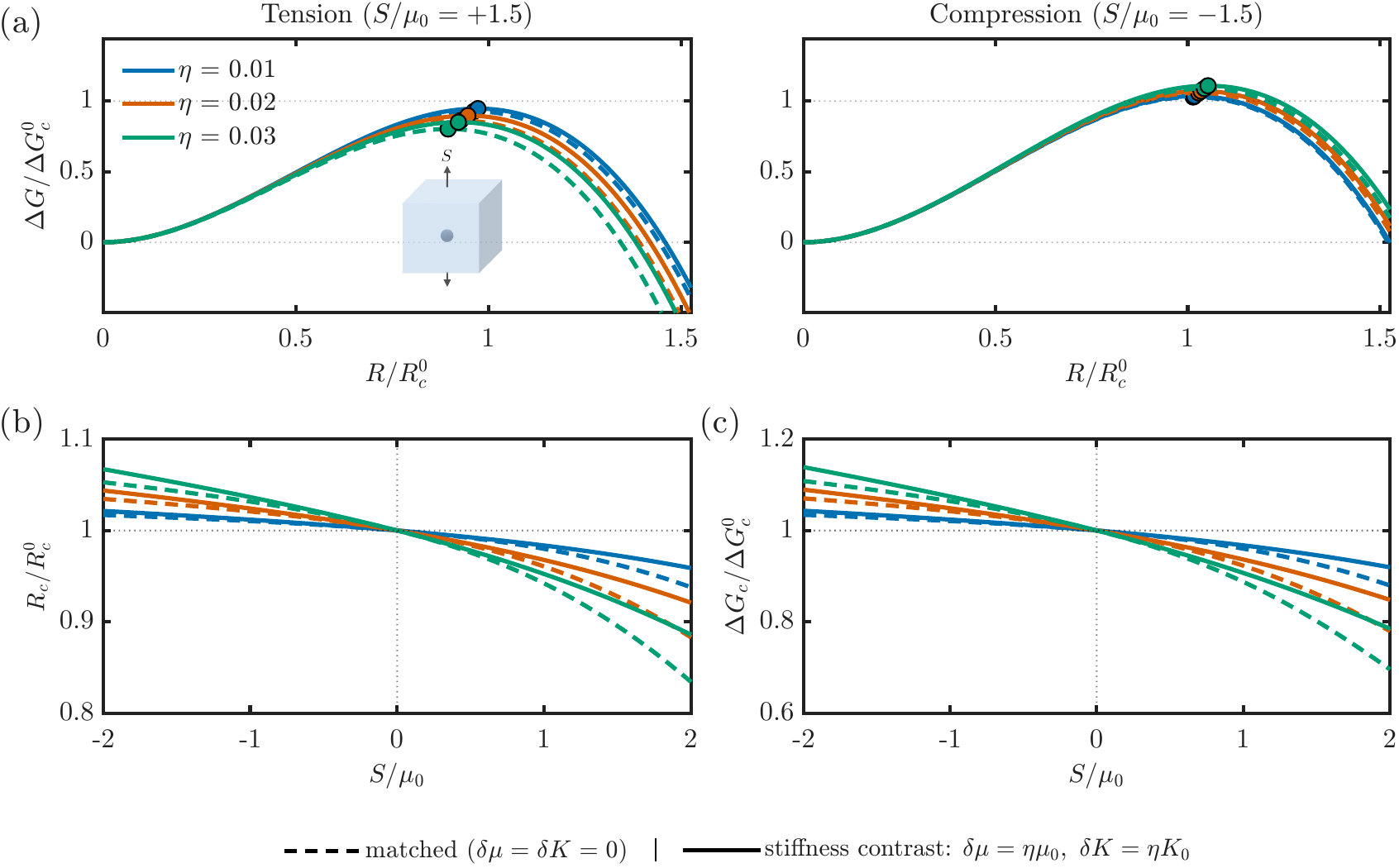}
    \caption{\textbf{Uniaxial stress modifies the CNT barrier through the finite-deformation prestress.}
(a) Normalized free-energy landscapes at $S/\mu_0=\pm1.5$; markers indicate critical points.
(b) Critical-radius ratio and (c) nucleation-barrier ratio versus $S/\mu_0$.
Results use $a=1$, $T_m-T=1\,\mathrm{K}$, and $\eta=\{0.01,0.02,0.03\}$.
Dashed curves denote matched stiffness; solid curves include
$\delta\mu=\eta\mu_0$ and $\delta K=\eta K_0$.}
    \label{fig:uni_G_a1}
\end{figure}%

\begin{figure}[H]
    \centering
    \includegraphics[width=\textwidth]{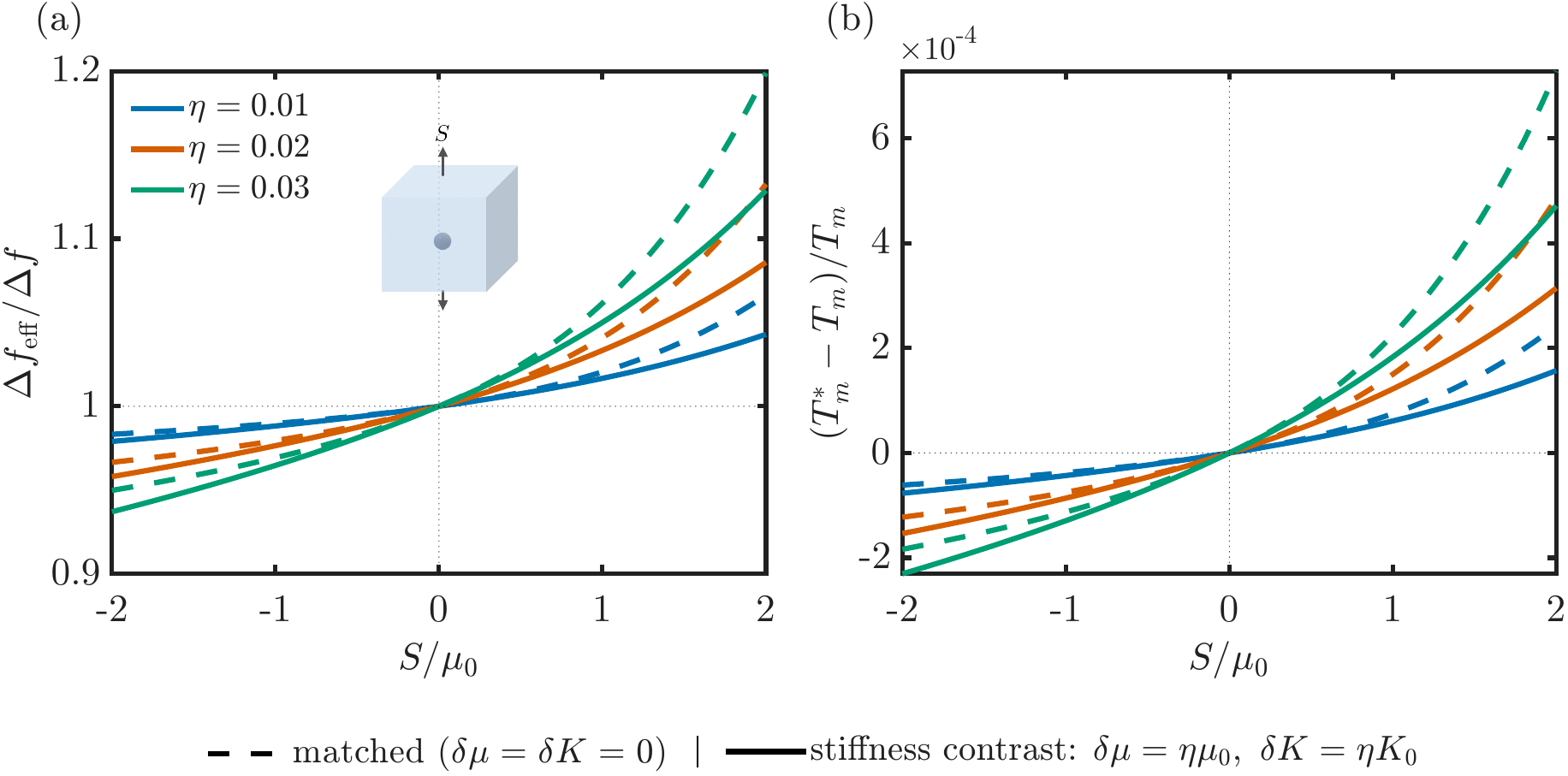}
    \caption{\textbf{Uniaxial stress changes the effective driving force for phase transformation.}
(a) Effective driving force normalized by the zero-stress thermal driving force, (shown for undercooling of $1$\,K) and
(b) relative critical-temperature shift versus normalized  stress. Results are shown for $a=1$ and $\eta=\{0.01,0.02,0.03\}$.
Dashed curves denote matched stiffness; solid curves include $\delta K=\eta K_0$ and $\delta \mu=\eta \mu_0$.
}

    \label{fig:uni_df_a1}
\end{figure}%

\subsection*{\textbf{Equibiaxial}}
\begin{figure}[H]
    \centering
    \includegraphics[width=\textwidth]{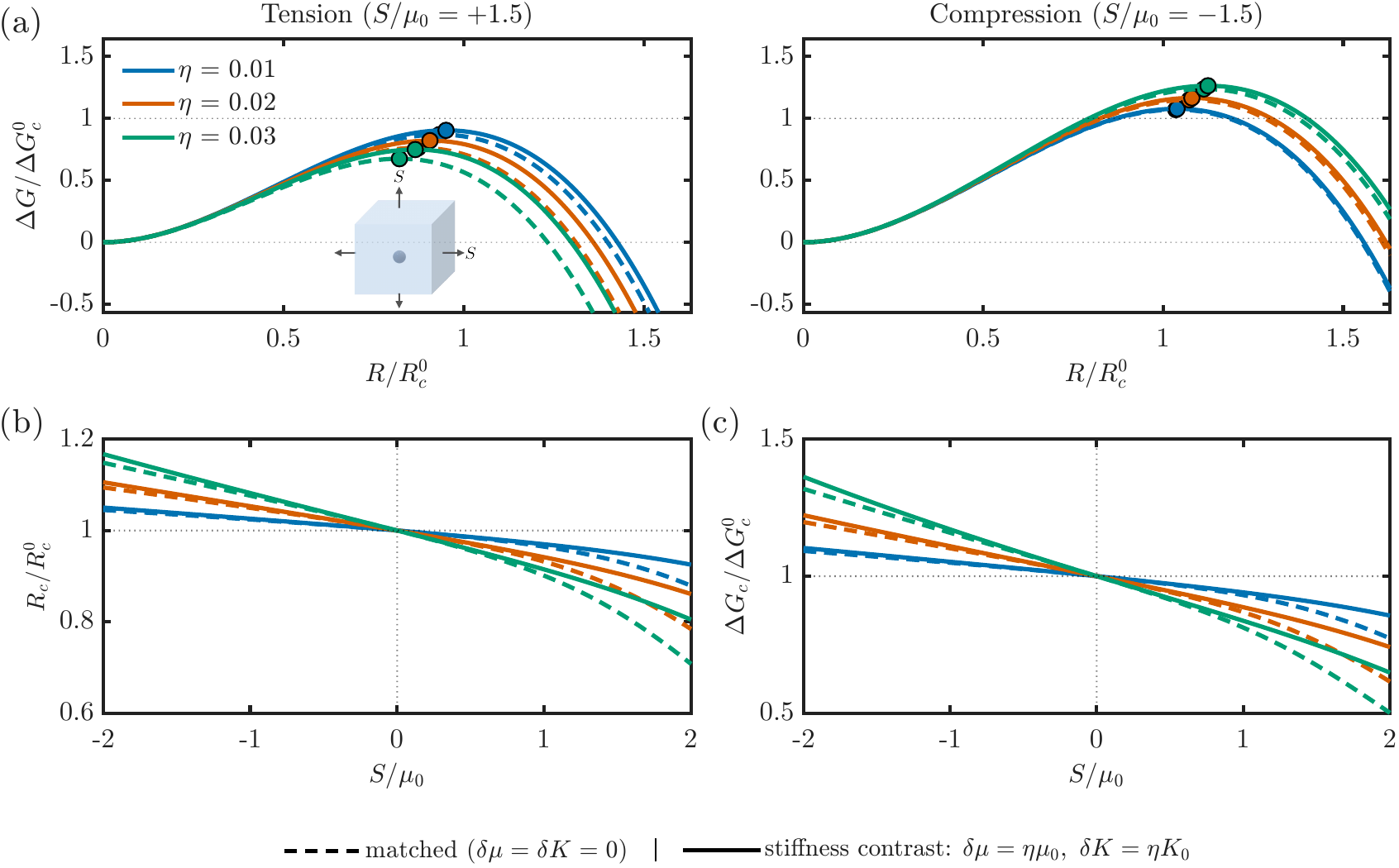}
     \caption{\textbf{Equibiaxial stress modifies the CNT barrier through the finite-deformation prestress.}
(a) Normalized free-energy landscapes at $S/\mu_0=\pm1.5$; markers indicate critical points.
(b) Critical-radius ratio and (c) nucleation-barrier ratio versus $S/\mu_0$.
Results use $a=1$, $T_m-T=1\,\mathrm{K}$, and $\eta=\{0.01,0.02,0.03\}$.
Dashed curves denote matched stiffness; solid curves include
$\delta\mu=\eta\mu_0$ and $\delta K=\eta K_0$.}
    \label{fig:bi_G_a1}
\end{figure}%

\begin{figure}[H]
    \centering
    \includegraphics[width=\textwidth]{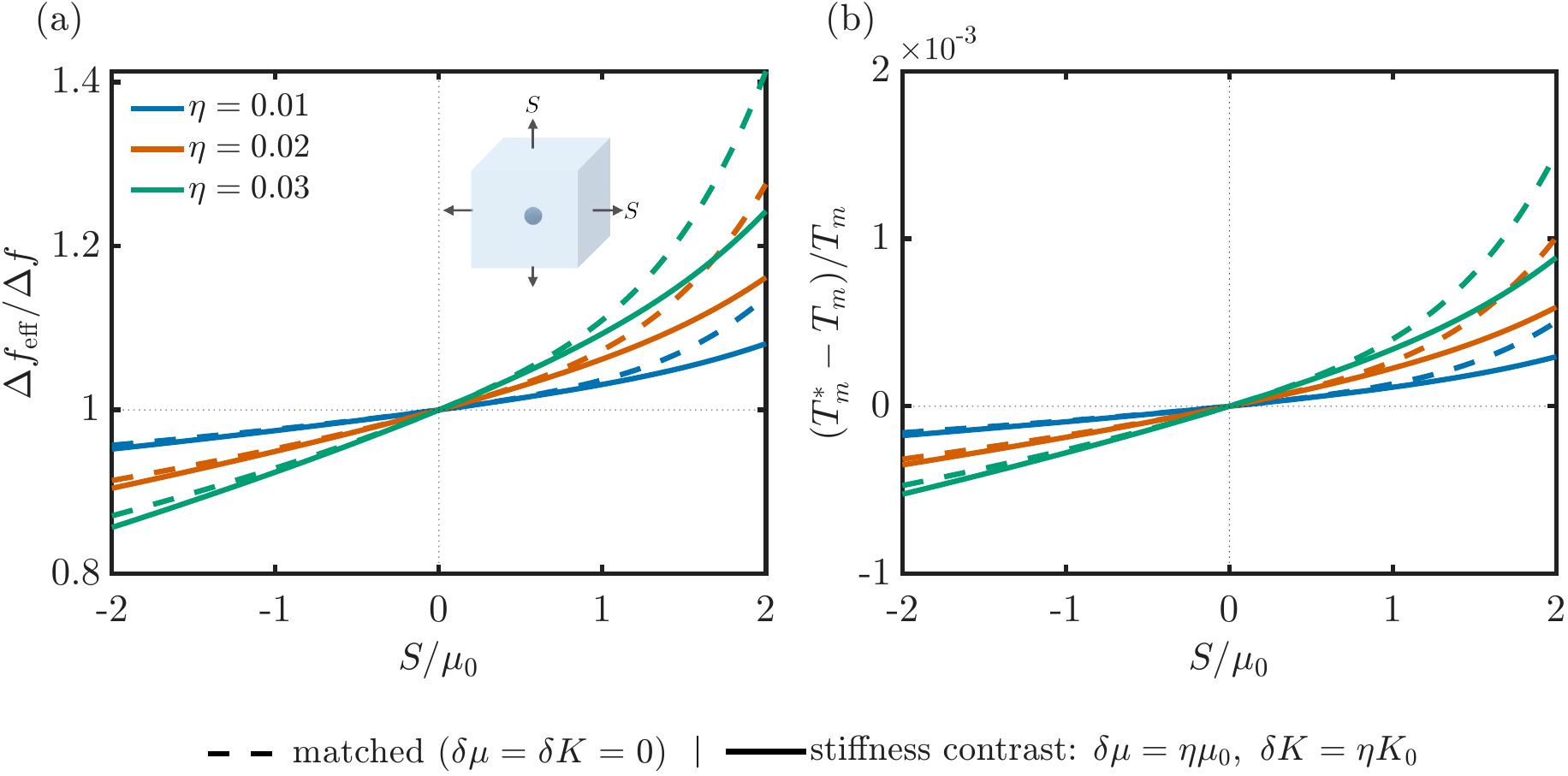}
    \caption{\textbf{Equibiaxial stress changes the effective driving force for phase transformation.}
(a) Effective driving force normalized by the zero-stress thermal driving force, (shown for undercooling of $1$\,K) and
(b) relative critical-temperature shift versus normalized  stress. Results are shown for $a=1$ and $\eta=\{0.01,0.02,0.03\}$.
Dashed curves denote matched stiffness; solid curves include $\delta K=\eta K_0$ and $\delta \mu=\eta \mu_0$.
}

    \label{fig:bi_df_a1}
\end{figure}%

\newpage

\paragraph{\textbf{II. Hydrostatic, Uniaxial, and Biaxial for a=1.002}}

\subsection*{\textbf{Hydrostatic}}
\begin{figure}[H]
    \centering
    \includegraphics[width=\textwidth]{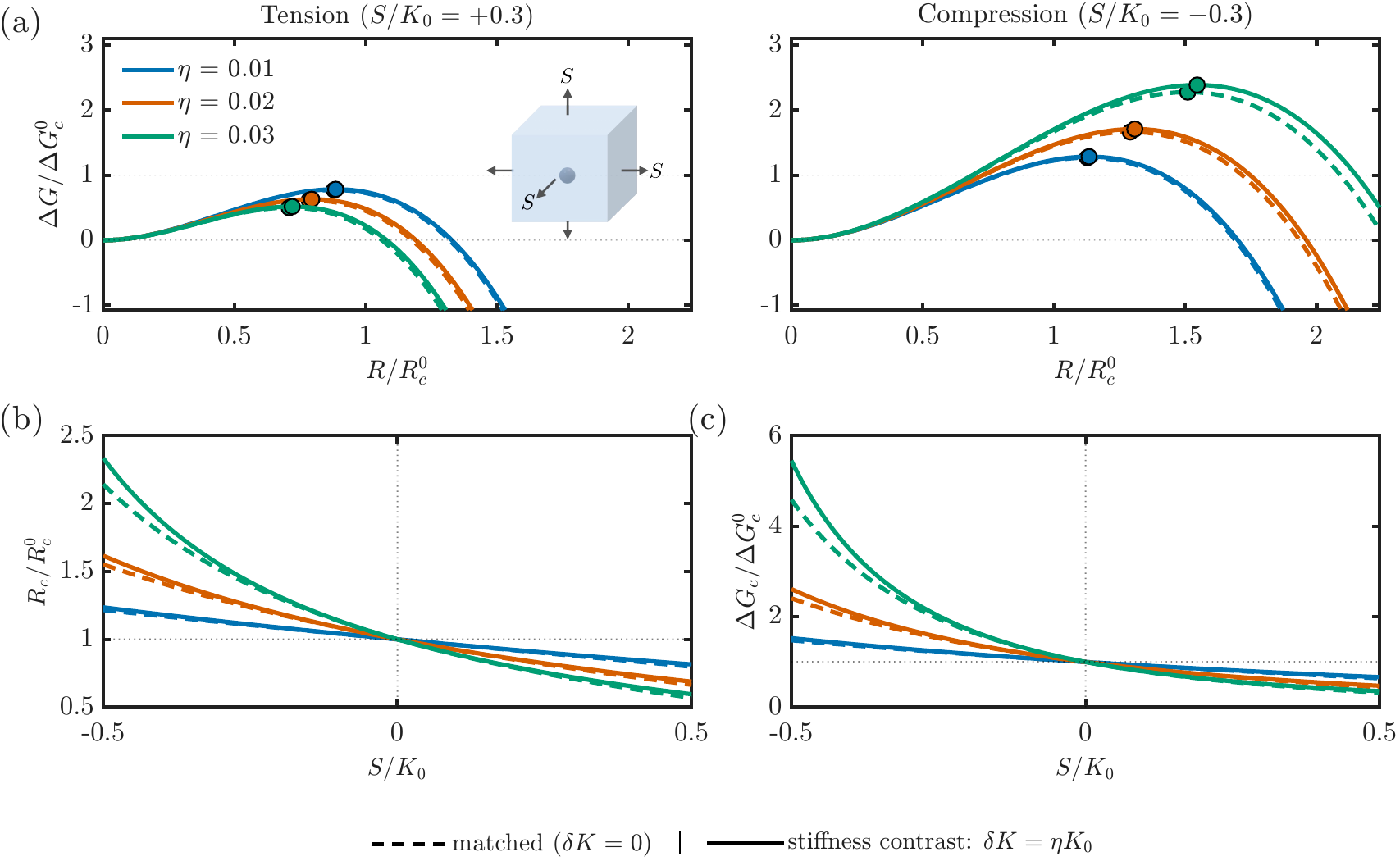}
  \caption{\textbf{Hydrostatic stress changes the CNT critical radius and barrier.}
(a) Normalized free-energy landscapes at $S/K_0=\pm0.3$; markers indicate critical points.
(b) Critical-radius ratio and (c) nucleation-barrier ratio versus $S/K_0$.
Results use $a=1.002$, $T_m-T=1\,\mathrm{K}$, and $\eta=\{0.01,0.02,0.03\}$.
Dashed curves denote matched stiffness; solid curves include $\delta K=\eta K_0$.}
   \label{hydro_G_a1p002}
\end{figure}%

\begin{figure}[H]
    \centering
    \includegraphics[width=\textwidth]{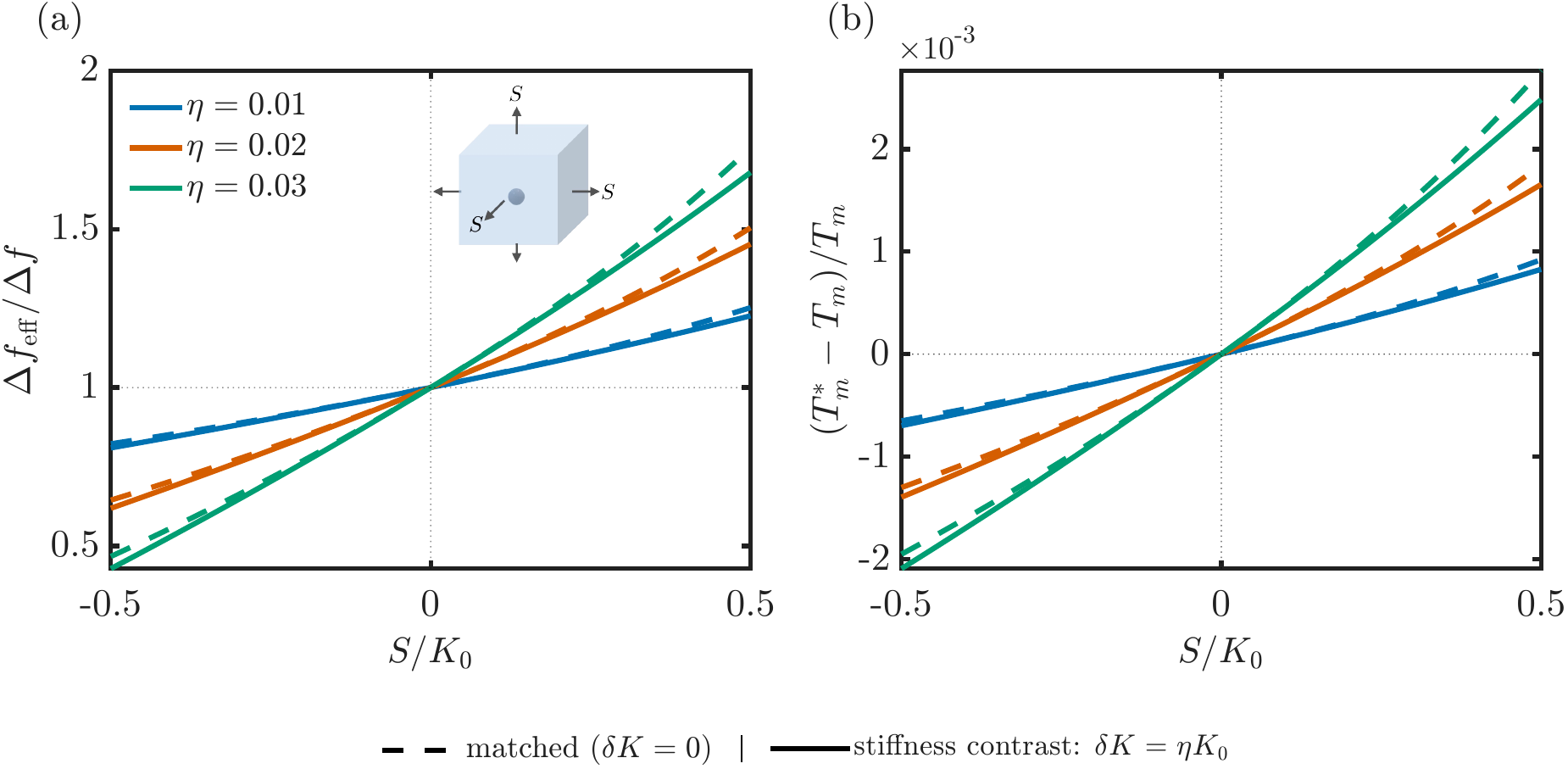}
  \caption{\textbf{Hydrostatic stress changes the effective driving force for phase transformation.}
(a) Effective driving force normalized by the zero-stress thermal driving force, (shown for undercooling of $1$\,K) and
(b) relative critical-temperature shift versus normalized stress. Results are shown for $a=1.002$ and $\eta=\{0.01,0.02,0.03\}$.
Dashed curves denote matched stiffness; solid curves include $\delta K=\eta K_0$.
For hydrostatic loading, the shear-modulus contrast does not contribute at first order.}
    \label{hydro_df_a1p002}
\end{figure}%

\subsection*{\textbf{Uniaxial}}
\begin{figure}[H]
    \centering
    \includegraphics[width=\textwidth]{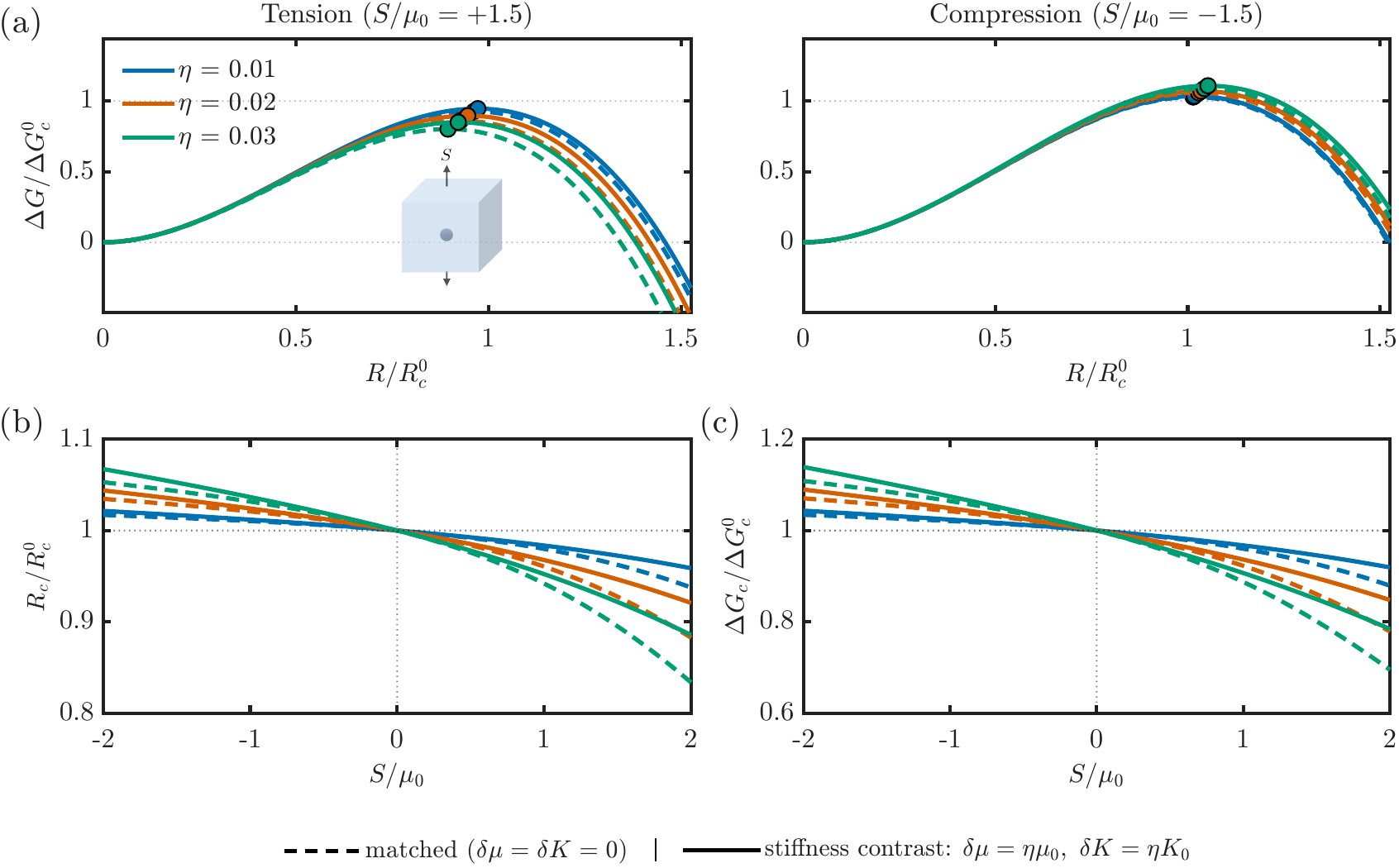}
 \caption{\textbf{Uniaxial stress modifies the CNT barrier through the finite-deformation prestress.}
(a) Normalized free-energy landscapes at $S/\mu_0=\pm1.5$; markers indicate critical points.
(b) Critical-radius ratio and (c) nucleation-barrier ratio versus $S/\mu_0$.
Results use $a=1.002$, $T_m-T=1\,\mathrm{K}$, and $\eta=\{0.01,0.02,0.03\}$.
Dashed curves denote matched stiffness; solid curves include
$\delta\mu=\eta\mu_0$ and $\delta K=\eta K_0$.}

    \label{uni_G_a1p002}
\end{figure}%

\begin{figure}[H]
    \centering
    \includegraphics[width=\textwidth]{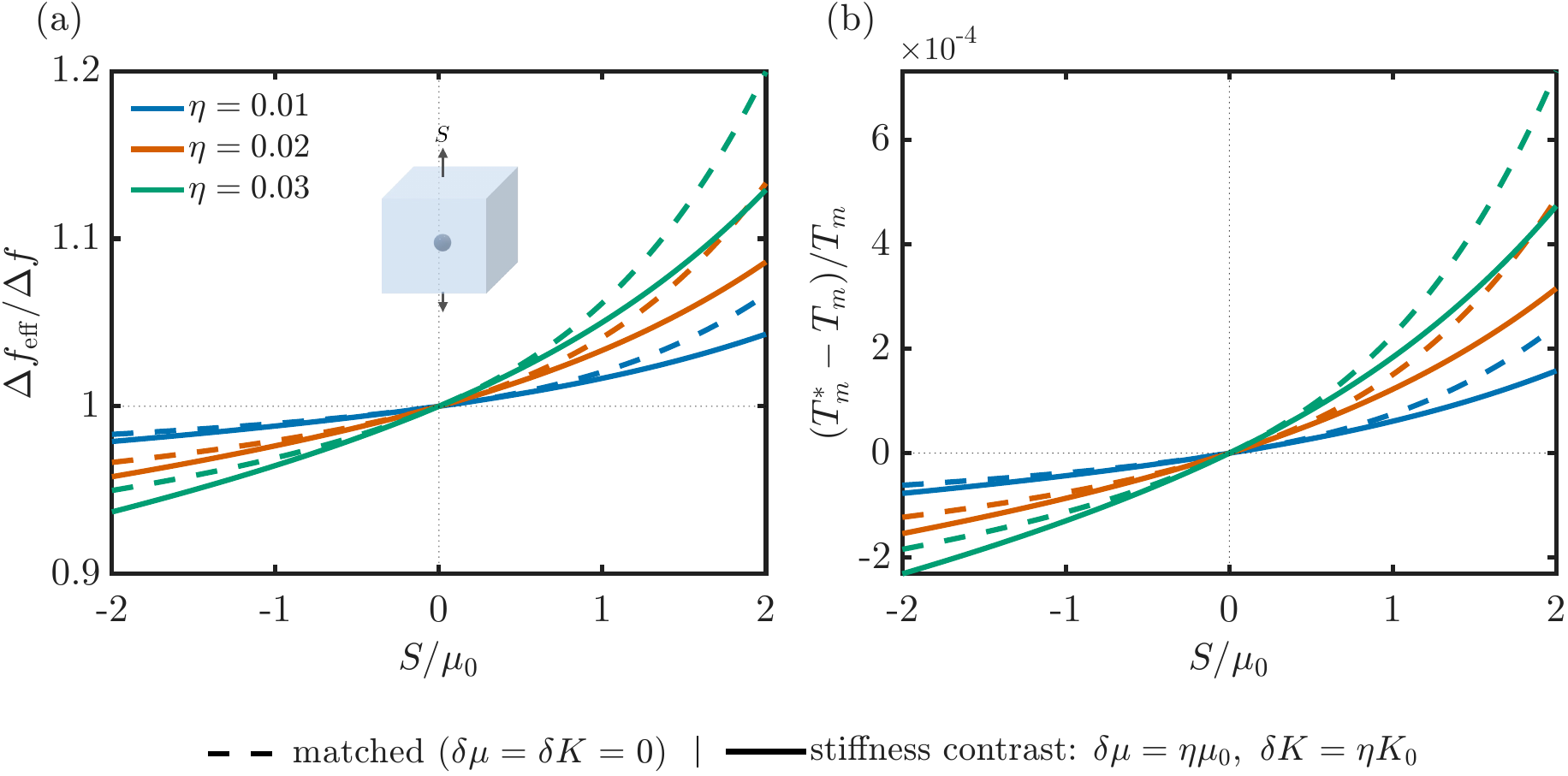}
  \caption{\textbf{Uniaxial stress changes the effective driving force for phase transformation.}
(a) Effective driving force normalized by the zero-stress thermal driving force, (shown for undercooling of $1$\,K) and
(b) relative critical-temperature shift versus normalized stress. Results are shown for $a=1.002$ and $\eta=\{0.01,0.02,0.03\}$.
Dashed curves denote matched stiffness; solid curves include $\delta K=\eta K_0$ and $\delta \mu=\eta \mu_0$.
}
  \label{uni_df_a1p002}
\end{figure}%

\subsection*{\textbf{Equibiaxial}}
\begin{figure}[H]
    \centering
    \includegraphics[width=\textwidth]{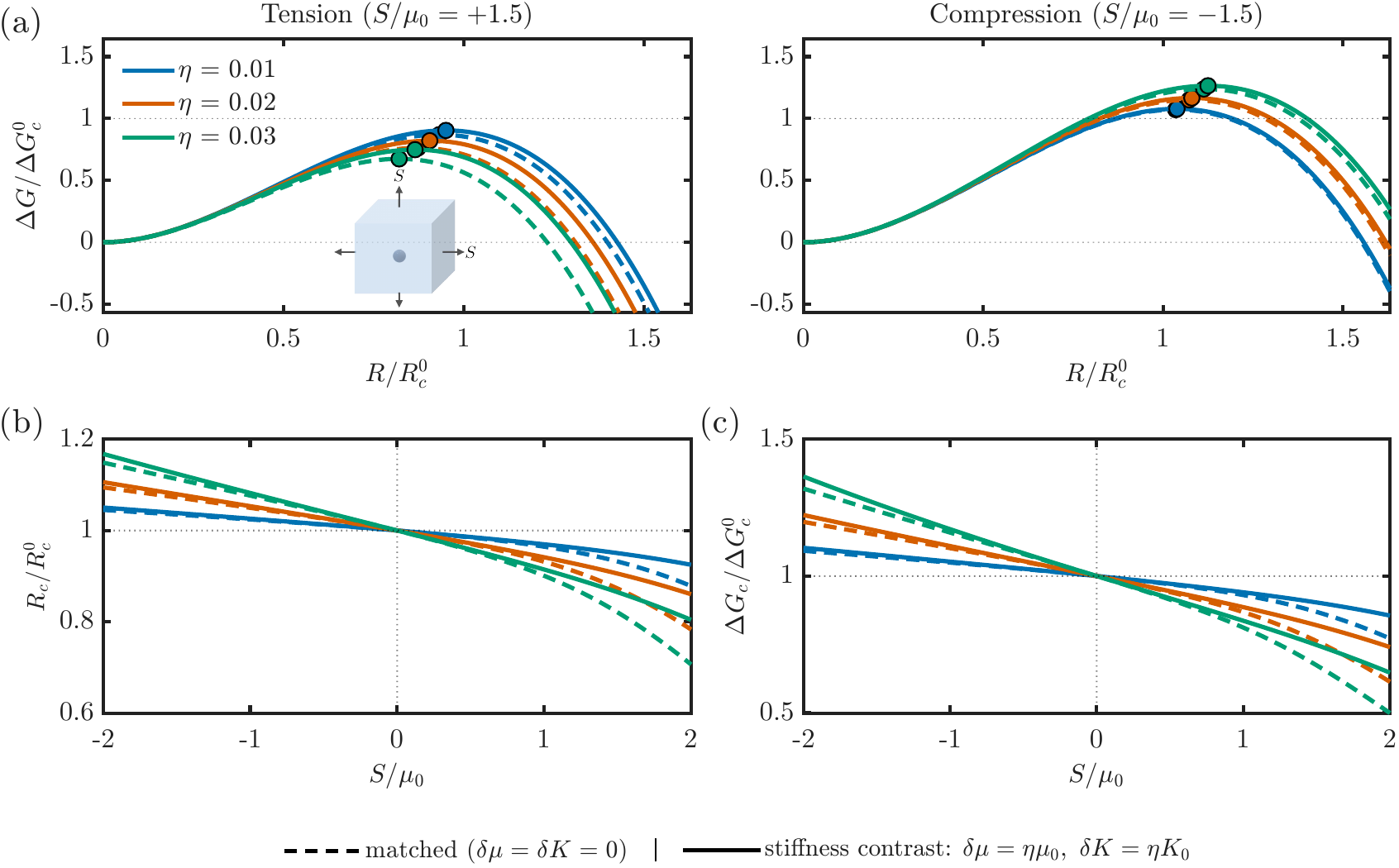}
 \caption{\textbf{Equibiaxial stress modifies the CNT barrier through the finite-deformation prestress.}
(a) Normalized free-energy landscapes at $S/\mu_0=\pm1.5$; markers indicate critical points.
(b) Critical-radius ratio and (c) nucleation-barrier ratio versus $S/\mu_0$.
Results use $a=1.002$, $T_m-T=1\,\mathrm{K}$, and $\eta=\{0.01,0.02,0.03\}$.
Dashed curves denote matched stiffness; solid curves include
$\delta\mu=\eta\mu_0$ and $\delta K=\eta K_0$.}
    \label{bi_G_a1p002}
\end{figure}%
\vspace{-3em}
\begin{figure}[H]
    \centering
    \includegraphics[width=\textwidth]{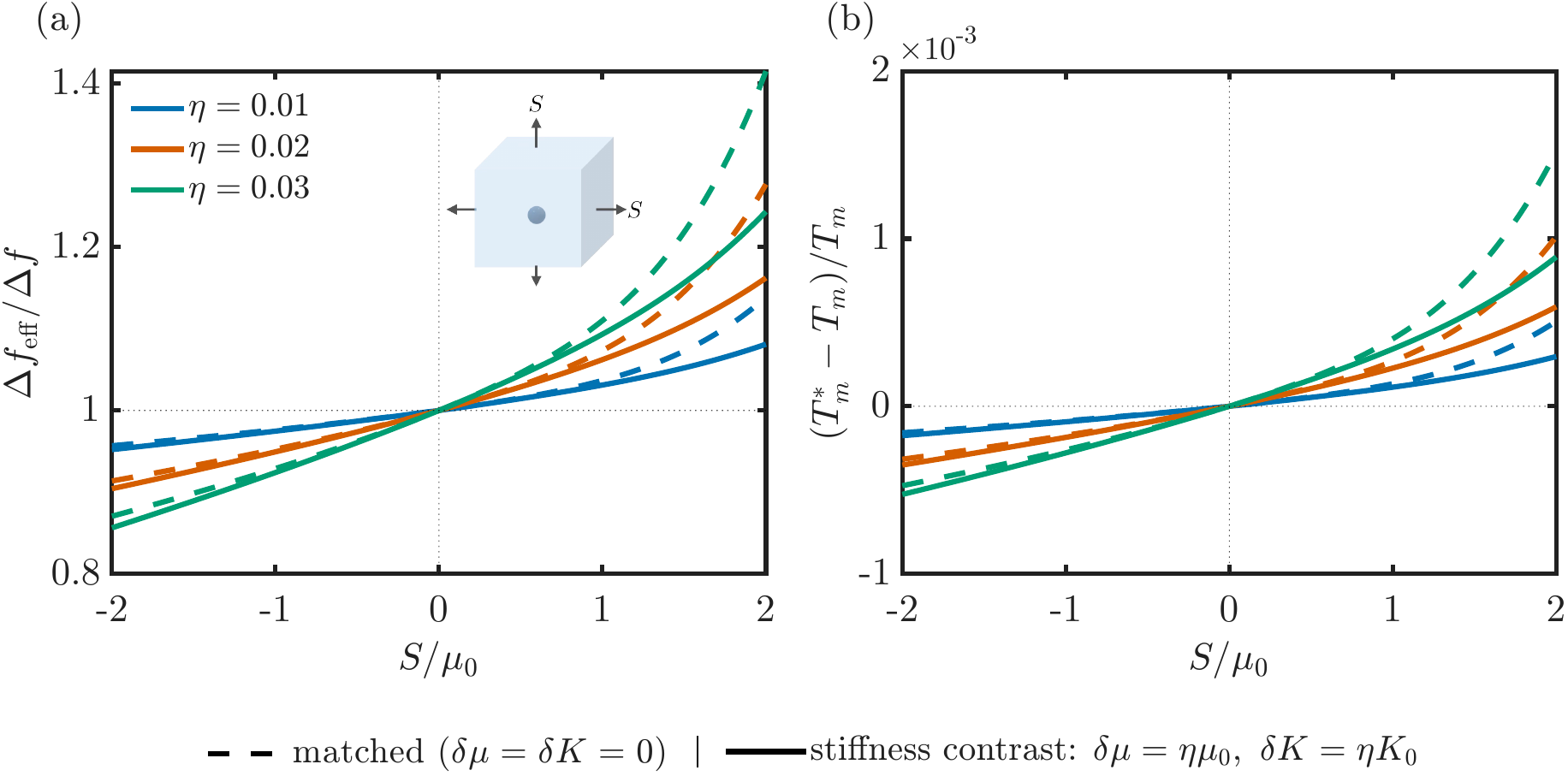}
 \caption{\textbf{Equibiaxial stress changes the effective driving force for phase transformation.}
(a) Effective driving force normalized by the zero-stress thermal driving force, (shown for undercooling of $1$\,K) and
(b) relative critical-temperature shift versus normalized  stress. Results are shown for $a=1.002$ and $\eta=\{0.01,0.02,0.03\}$.
Dashed curves denote matched stiffness; solid curves include $\delta K=\eta K_0$ and $\delta \mu=\eta \mu_0$.
}
    \label{bi_df_a1p002}
\end{figure}%

\newpage
{\color{myblue}
\section{Direct numerical validation of the first-order energy}\label{app:FE}

The first-order estimate~\eqref{eq:DeltaPi_NH_mixed_main} was validated by directly solving the nonlinear elasticity boundary-value problem for the compressible neo-Hookean solid~\eqref{eq:NH_energy_main}, with $K_0=10\,$MPa, $\mu_0=1\,$MPa, $\delta K = \eta K_0$, $\delta \mu = \eta \mu_0$ and $a=1$.

\subsection*{\textbf{Uniaxial and equibiaxial loading: 3-D Finite Element Simulation in FEniCS}}
For non-hydrostatic loading the fields are not radially symmetric and the problem is solved in three dimensions. 
The spherical nucleus of radius $A$ is embedded in a \emph{cubic} matrix of half-edge $B=50A$; one octant is modeled, with symmetry boundary conditions. 
The remote stress is applied as a dead-load traction $\mathbf t_0=\Pz\mathbf N$ on the outer faces. 
The nucleus differs from the matrix only through $\eta$ (eigenstrain $a(1+\eta)\id$ and moduli $\mu_0+\eta\mu_0$, $K_0+\eta K_0$ inside), so at $\eta=0$ the body is homogeneous. The mesh is graded tetrahedra and the displacement uses quadratic (P2) Lagrange interpolation.
Equilibrium is solved by Newton's method (PETSc SNES) with a direct (MUMPS) linear solver, with the load applied incrementally. 
The potential energy difference $\Delta\Pi=\Pi(\eta)-\Pi(0)$ is evaluated on the same mesh.

Two exact checks underpin these results. 
First, at $\eta=0$ the potential energy of the homogeneous body calculated analytically and through finite-element analysis differ by $\le3\times10^{-14}$ (relative) at every stress and loading, verifying the energy functional and the dead-load term; similarly the base stretches also match the analytic homogeneous stretches of~\ref{app:diagonal_states}. 
Second, the result is independent of the outer domain: repeating the computation on a \emph{spherical} matrix of radius $B$ changes $\Delta\Pi/V_i$ by at most $5\times10^{-5}$ (relative) over all stress levels and both loadings---three orders of magnitude below the $O(\eta^2)$ accommodation---confirming that the truncation $B/A=50$ and the outer-boundary shape are immaterial. 

Figures~\ref{fig:FEuni}(a) and~\ref{fig:FEbi}(a) compare the asymptotic first-order prediction of $\Delta\Pi/V_i$  as noted in Table~\ref{tab:DeltaPi_PK_cases} (original contribution of this work) with finite element predictions (considered as ``truth''); panel (b) in both figures shows the relative error in the first order estimate when compared to the finite element simulations.
Similar to the case of hydrostatic loading, the deviation is the $O(\eta^2)$ accommodation, growing near $S=0$ where the first-order term vanishes (we have omitted connecting lines across $S=0$ in panel (b) for this reason to avoid confusion). 
To further verify the veracity of the result, we carry out a sweep over $\eta \in \{ 0.005, 0.01, 0.02, 0.03 \}$ for $S/\mu_0=1$ for matched stiffness as well as contrasting stiffness cases. 
These results, presented in panel (c) of each figure, confirm that the error indeed scales as $O(\eta^2)$.
Thus, the asymptotic approximation keeps getting better as $\eta \rightarrow 0$.

\begin{figure}[H]
\centering
\includegraphics[width=\textwidth]{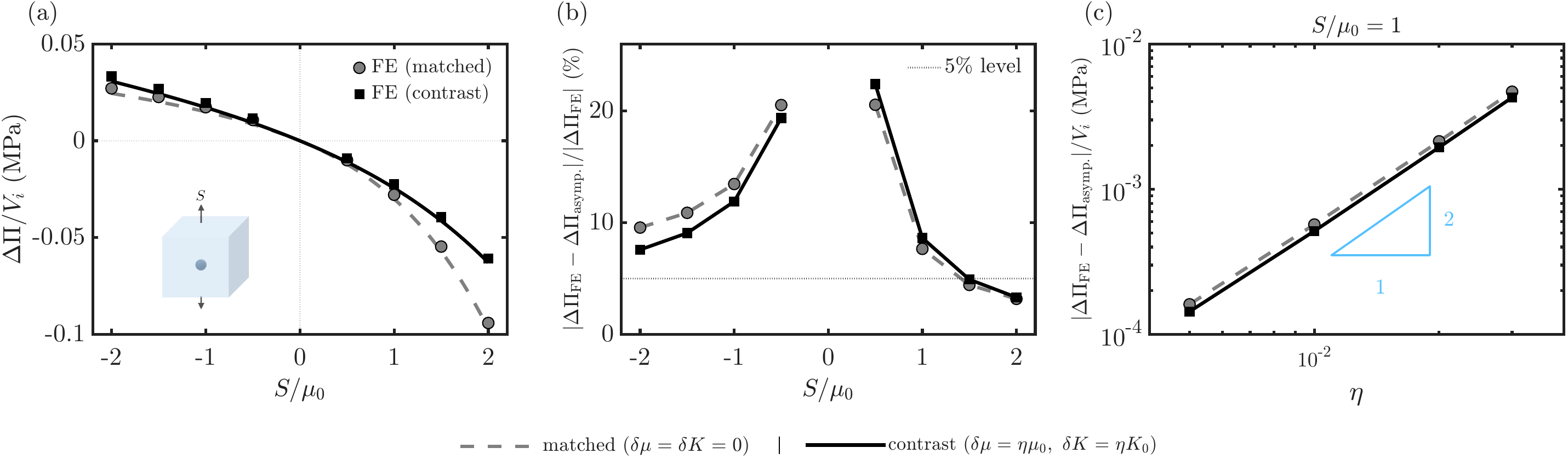}
\caption{\myblue{\textbf{Uniaxial validation (cubic matrix domain).} (a) $\Delta\Pi/V_i$ versus $S/\mu_0$: first-order asymptotic (lines, Table~\ref{tab:DeltaPi_PK_cases}) and three-dimensional finite-element solution (markers), for matched and stiffness-contrast cases. (b) Relative deviation of the first-order formula; the dashed line marks the $5\%$ level. (c) Absolute error $|\Delta\Pi_{\rm FE}-\Delta\Pi_{\rm asymp.}|/V_i$ versus $\eta$ at fixed stress $S/\mu_0=1$: the first-order estimate converges to the direct solution as $O(\eta^2)$ (slope-2 reference triangle).  Dashed gray curves show matched-stiffness and solid black curves show stiffness-contrast cases.}}
\label{fig:FEuni}
\end{figure}

\begin{figure}[H]
\centering
\includegraphics[width=\textwidth]{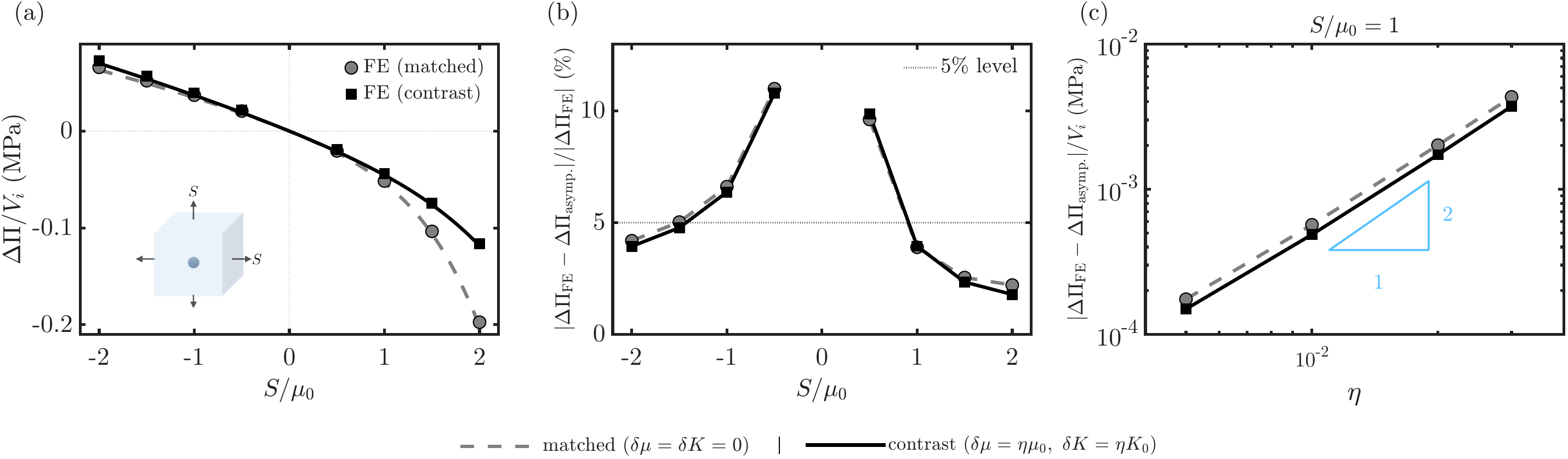}
\caption{\myblue{\textbf{Equibiaxial validation (cubic matrix domain)}. Plotted as in Fig.~\ref{fig:FEuni}.}}
\label{fig:FEbi}
\end{figure}
}

\newpage
\bibliographystyle{elsarticle-num}
\bibliography{StressNucleation_refs}
\end{document}